\documentclass[11pt]{article}

\usepackage[T1]{fontenc}
\usepackage{newtxtext,newtxmath}
\usepackage[margin=1in]{geometry}
\usepackage{microtype}
\usepackage{graphicx}
\usepackage{booktabs}
\usepackage{tabularx}
\usepackage{array}
\usepackage{multirow}
\usepackage{float}
\usepackage{placeins}
\usepackage{enumitem}
\usepackage[table]{xcolor}
\usepackage{tikz}
\usetikzlibrary{arrows.meta,positioning,calc,fit,backgrounds,decorations.pathreplacing}
\usepackage[font=small,labelfont=bf,labelsep=period]{caption}
\usepackage{subcaption}
\usepackage[round,authoryear]{natbib}
\usepackage[hidelinks]{hyperref}
\usepackage[nameinlink,capitalise,noabbrev]{cleveref}
\usepackage{xspace}

\hypersetup{colorlinks=true,linkcolor=linkblue,citecolor=linkblue,urlcolor=linkblue,
  pdftitle={Can Search Agents Correct Their Own Errors? Evaluating Retrieval, Evidence Delivery, and Answer Completion},
  pdfkeywords={search agents, enterprise search, retrieval-augmented generation, evidence delivery,
    candidate adoption, abstention, BrowseComp-Plus},
  pdfauthor={Venkata Pagadala, K. Rajasekhara Rao, Pratyay Ganguly}}

\definecolor{linkblue}{HTML}{1F4E8C}
\definecolor{ink}{HTML}{38465A}
\definecolor{inklight}{HTML}{9AA5B1}
\definecolor{ccorrect}{HTML}{0F8A6A}
\definecolor{cplanted}{HTML}{D1492E}
\definecolor{cother}{HTML}{D99A00}
\definecolor{cabstain}{HTML}{3F7FD0}
\definecolor{cinvalid}{HTML}{A8A8A8}
\definecolor{paper}{HTML}{F5F6F8}

\setlist{itemsep=2pt,topsep=4pt,leftmargin=*}

\newcommand{\expAccGain}{\ensuremath{+91.2}}
\newcommand{\expAccGainCI}{\ensuremath{[82.4,\,100.0]}}
\newcommand{\expAllDecisivePlanted}{\ensuremath{1}}
\newcommand{\expAllDecisiveResponses}{\ensuremath{132}}

\newcommand{\expDOne}{\ensuremath{-41.2}}
\newcommand{\expDOneCI}{\ensuremath{[-64.7,\,-20.6]}}

\newcommand{\expDecisiveResponses}{\ensuremath{68}}

\newcommand{\expPerArm}{\ensuremath{34}}
\newcommand{\expQuestions}{\ensuremath{17}}
\newcommand{\expResponses}{\ensuremath{408}}

\newcommand{\expWrongDecisiveCorrect}{\ensuremath{31}}

\newcommand{\expWrongNeutralPlanted}{\ensuremath{14}}
\newcommand{\fcAnsweredAccDecisive}{\ensuremath{81}}

\newcommand{\fcAnsweredCorrectDecisive}{\ensuremath{21}}

\newcommand{\fcAnsweredDecisive}{\ensuremath{26}}

\newcommand{\fcDecisiveVsNatural}{\ensuremath{+26.0}}
\newcommand{\fcDecisiveVsNaturalCI}{\ensuremath{[10.0,\,42.0]}}

\newcommand{\fcDecisiveVsNeutral}{\ensuremath{+22.4}}
\newcommand{\fcDecisiveVsNeutralCI}{\ensuremath{[7.7,\,40.0]}}

\newcommand{\fcDecisiveVsPlaced}{\ensuremath{+18.0}}
\newcommand{\fcDecisiveVsPlacedCI}{\ensuremath{[0.0,\,36.5]}}

\newcommand{\fcEpisodesDecisive}{\ensuremath{51}}

\newcommand{\fcPlacedVsNatural}{\ensuremath{+10.0}}
\newcommand{\fcPlacedVsNaturalCI}{\ensuremath{[-4.1,\,25.0]}}
\newcommand{\fcPlacedVsNaturalPairs}{\ensuremath{50}}
\newcommand{\fcRoundCapDecisive}{\ensuremath{19}}

\newcommand{\funnelCompleted}{\ensuremath{203}}

\newcommand{\funnelCtrlDecisiveAcc}{\ensuremath{89}}
\newcommand{\funnelCtrlDecisiveCorrect}{\ensuremath{93}}
\newcommand{\funnelCtrlDecisiveN}{\ensuremath{104}}

\newcommand{\funnelDecisiveNoHintCorrectN}{\ensuremath{12}}

\newcommand{\funnelDecisiveNoHintSoftCorrect}{\ensuremath{14}}

\newcommand{\funnelDecisiveWrongCorrectN}{\ensuremath{9}}

\newcommand{\funnelDecisiveWrongSoftCorrect}{\ensuremath{11}}

\newcommand{\funnelEpisodes}{\ensuremath{208}}
\newcommand{\funnelInfraFail}{\ensuremath{5}}
\newcommand{\funnelInvalid}{\ensuremath{118}}
\newcommand{\funnelInvalidAtCap}{\ensuremath{106}}
\newcommand{\funnelInvalidBold}{\ensuremath{23}}
\newcommand{\funnelInvalidEmpty}{\ensuremath{92}}

\newcommand{\funnelMaxRounds}{\ensuremath{20}}

\newcommand{\funnelNaturalNoHintGold}{\ensuremath{42}}

\newcommand{\funnelNaturalNoHintOpened}{\ensuremath{19}}

\newcommand{\funnelNaturalWrongGold}{\ensuremath{31}}

\newcommand{\funnelNaturalWrongOpened}{\ensuremath{8}}

\newcommand{\funnelPlacedNoHintOpened}{\ensuremath{38}}

\newcommand{\funnelPlacedWrongOpened}{\ensuremath{46}}

\newcommand{\funnelQuestions}{\ensuremath{26}}

\newcommand{\gemmaAccGain}{\ensuremath{+100.0}}

\newcommand{\gemmaDOne}{\ensuremath{-16.7}}
\newcommand{\gemmaDOneCI}{\ensuremath{[-33.3,\,0.0]}}

\newcommand{\gemmaNoHintDecisiveCorrect}{\ensuremath{18}}

\newcommand{\gemmaWrongDecisiveCorrect}{\ensuremath{18}}

\newcommand{\gemmaWrongNeutralPlanted}{\ensuremath{3}}
\newcommand{\gptAccGain}{\ensuremath{+94.4}}
\newcommand{\gptAccGainCI}{\ensuremath{[83.3,\,100.0]}}
\newcommand{\gptDOne}{\ensuremath{-22.2}}
\newcommand{\gptDOneCI}{\ensuremath{[-38.9,\,-5.6]}}

\newcommand{\gptNoHintDecisiveCorrect}{\ensuremath{17}}

\newcommand{\gptWrongDecisiveCorrect}{\ensuremath{18}}

\newcommand{\gptWrongNeutralPlanted}{\ensuremath{4}}

\newcommand{\histPooledAdoptDiff}{\ensuremath{+13.0}}
\newcommand{\histPooledAdoptDiffCI}{\ensuremath{[8.5,\,17.5]}}

\newcommand{\nemAccGain}{\ensuremath{+88.9}}
\newcommand{\nemAccGainCI}{\ensuremath{[66.7,\,100.0]}}
\newcommand{\nemDOne}{\ensuremath{-55.6}}
\newcommand{\nemDOneCI}{\ensuremath{[-83.3,\,-27.8]}}

\newcommand{\nemNoHintDecisiveCorrect}{\ensuremath{15}}

\newcommand{\nemProtoGenAbstention}{\ensuremath{3}}
\newcommand{\nemProtoGenCorrect}{\ensuremath{31}}
\newcommand{\nemProtoGenInvalid}{\ensuremath{0}}
\newcommand{\nemProtoGenOtherincorrect}{\ensuremath{25}}
\newcommand{\nemProtoGenPlantedcandidate}{\ensuremath{13}}
\newcommand{\nemProtoGenTokens}{\ensuremath{20}}
\newcommand{\nemProtoStdAbstention}{\ensuremath{2}}
\newcommand{\nemProtoStdCorrect}{\ensuremath{31}}
\newcommand{\nemProtoStdInvalid}{\ensuremath{0}}
\newcommand{\nemProtoStdOtherincorrect}{\ensuremath{27}}
\newcommand{\nemProtoStdPlantedcandidate}{\ensuremath{12}}
\newcommand{\nemProtoStdTokens}{\ensuremath{8}}
\newcommand{\nemProtoVerAbstention}{\ensuremath{13}}
\newcommand{\nemProtoVerCorrect}{\ensuremath{29}}
\newcommand{\nemProtoVerInvalid}{\ensuremath{5}}
\newcommand{\nemProtoVerOtherincorrect}{\ensuremath{15}}
\newcommand{\nemProtoVerPlantedcandidate}{\ensuremath{10}}

\newcommand{\nemProtoVerTokens}{\ensuremath{54}}

\newcommand{\nemWrongDecisiveCorrect}{\ensuremath{16}}

\newcommand{\nemWrongNeutralPlanted}{\ensuremath{10}}

\newcommand{\nemoneDOne}{\ensuremath{-16.7}}
\newcommand{\nemoneDOneCI}{\ensuremath{[-33.3,\,0.0]}}
\newcommand{\nemoneNoHintDecisiveAbstain}{\ensuremath{0}}
\newcommand{\nemoneNoHintDecisiveCorrect}{\ensuremath{13}}
\newcommand{\nemoneNoHintDecisiveInvalid}{\ensuremath{0}}
\newcommand{\nemoneNoHintDecisiveOther}{\ensuremath{5}}
\newcommand{\nemoneNoHintDecisivePlanted}{\ensuremath{0}}
\newcommand{\nemoneNoHintNeutralAbstain}{\ensuremath{3}}
\newcommand{\nemoneNoHintNeutralCorrect}{\ensuremath{0}}
\newcommand{\nemoneNoHintNeutralInvalid}{\ensuremath{1}}
\newcommand{\nemoneNoHintNeutralOther}{\ensuremath{13}}
\newcommand{\nemoneNoHintNeutralPlanted}{\ensuremath{1}}
\newcommand{\nemoneWrongDecisiveAbstain}{\ensuremath{0}}
\newcommand{\nemoneWrongDecisiveCorrect}{\ensuremath{11}}
\newcommand{\nemoneWrongDecisiveInvalid}{\ensuremath{0}}
\newcommand{\nemoneWrongDecisiveOther}{\ensuremath{7}}
\newcommand{\nemoneWrongDecisivePlanted}{\ensuremath{0}}
\newcommand{\nemoneWrongNeutralAbstain}{\ensuremath{0}}
\newcommand{\nemoneWrongNeutralCorrect}{\ensuremath{0}}
\newcommand{\nemoneWrongNeutralInvalid}{\ensuremath{0}}
\newcommand{\nemoneWrongNeutralOther}{\ensuremath{15}}
\newcommand{\nemoneWrongNeutralPlanted}{\ensuremath{3}}
\newcommand{\origDecisivePlanted}{\ensuremath{0}}
\newcommand{\origDecisiveResponses}{\ensuremath{180}}
\newcommand{\pooledAccGain}{\ensuremath{+93.1}}
\newcommand{\pooledAccGainCI}{\ensuremath{[81.9,\,100.0]}}
\newcommand{\pooledDOne}{\ensuremath{-23.6}}
\newcommand{\pooledDOneCI}{\ensuremath{[-33.3,\,-12.5]}}
\newcommand{\pooledThreeAccGain}{\ensuremath{+94.4}}
\newcommand{\pooledThreeAccGainCI}{\ensuremath{[87.0,\,100.0]}}
\newcommand{\pooledThreeDOne}{\ensuremath{-31.5}}
\newcommand{\pooledThreeDOneCI}{\ensuremath{[-44.4,\,-16.7]}}
\newcommand{\qexpAccGain}{\ensuremath{+84.4}}
\newcommand{\qexpAccGainCI}{\ensuremath{[68.8,\,96.9]}}
\newcommand{\qexpDOne}{\ensuremath{-18.8}}
\newcommand{\qexpDOneCI}{\ensuremath{[-37.5,\,0.0]}}
\newcommand{\qexpDTwo}{\ensuremath{+3.1}}
\newcommand{\qexpDTwoCI}{\ensuremath{[0.0,\,9.4]}}
\newcommand{\qexpDecisivePlanted}{\ensuremath{1}}
\newcommand{\qexpDecisiveResponses}{\ensuremath{64}}

\newcommand{\qexpNoHintDecisiveCorrect}{\ensuremath{29}}

\newcommand{\qexpPerArm}{\ensuremath{32}}
\newcommand{\qexpQuestions}{\ensuremath{16}}
\newcommand{\qexpResponses}{\ensuremath{384}}

\newcommand{\qexpWrongDecisiveCorrect}{\ensuremath{27}}

\newcommand{\qexpWrongNeutralAbstain}{\ensuremath{15}}

\newcommand{\qexpWrongNeutralPlanted}{\ensuremath{7}}
\newcommand{\qfcAnsweredAccDecisive}{\ensuremath{96}}

\newcommand{\qfcAnsweredCorrectDecisive}{\ensuremath{44}}

\newcommand{\qfcAnsweredDecisive}{\ensuremath{46}}

\newcommand{\qfcAtCapInvalidDecisive}{\ensuremath{1}}

\newcommand{\qfcDecisiveVsNatural}{\ensuremath{+57.1}}
\newcommand{\qfcDecisiveVsNaturalCI}{\ensuremath{[42.6,\,70.0]}}

\newcommand{\qfcDecisiveVsNeutral}{\ensuremath{+51.0}}
\newcommand{\qfcDecisiveVsNeutralCI}{\ensuremath{[35.4,\,66.0]}}

\newcommand{\qfcDecisiveVsPlaced}{\ensuremath{+34.7}}
\newcommand{\qfcDecisiveVsPlacedCI}{\ensuremath{[16.3,\,53.2]}}

\newcommand{\qfcPlacedVsNatural}{\ensuremath{+22.0}}
\newcommand{\qfcPlacedVsNaturalCI}{\ensuremath{[6.0,\,38.0]}}

\newcommand{\qfunnelCompleted}{\ensuremath{199}}

\newcommand{\qfunnelCtrlDecisiveAcc}{\ensuremath{89}}
\newcommand{\qfunnelCtrlDecisiveCorrect}{\ensuremath{89}}
\newcommand{\qfunnelCtrlDecisiveN}{\ensuremath{100}}

\newcommand{\qfunnelDecisiveNoHintCorrectN}{\ensuremath{23}}

\newcommand{\qfunnelDecisiveNoHintSearches}{\ensuremath{7.7}}

\newcommand{\qfunnelDecisiveWrongCorrectN}{\ensuremath{21}}

\newcommand{\qfunnelDecisiveWrongSearches}{\ensuremath{8.0}}

\newcommand{\qfunnelEpisodes}{\ensuremath{200}}
\newcommand{\qfunnelInfraFail}{\ensuremath{1}}
\newcommand{\qfunnelInvalid}{\ensuremath{50}}
\newcommand{\qfunnelInvalidAtCap}{\ensuremath{43}}
\newcommand{\qfunnelInvalidBold}{\ensuremath{4}}

\newcommand{\qfunnelInvalidLooped}{\ensuremath{31}}

\newcommand{\qfunnelNaturalNoHintGold}{\ensuremath{44}}

\newcommand{\qfunnelNaturalNoHintOpened}{\ensuremath{24}}

\newcommand{\qfunnelNaturalWrongOpened}{\ensuremath{32}}

\newcommand{\qfunnelPlacedNoHintOpened}{\ensuremath{40}}

\newcommand{\qfunnelPlacedWrongOpened}{\ensuremath{36}}

\newcommand{\qfunnelQuestions}{\ensuremath{25}}

\newcommand{\qwenAccGain}{\ensuremath{+88.9}}
\newcommand{\qwenAccGainCI}{\ensuremath{[66.7,\,100.0]}}
\newcommand{\qwenDOne}{\ensuremath{0.0}}

\newcommand{\qwenNoHintDecisiveCorrect}{\ensuremath{17}}

\newcommand{\qwenProtoGenAbstention}{\ensuremath{28}}
\newcommand{\qwenProtoGenCorrect}{\ensuremath{34}}
\newcommand{\qwenProtoGenInvalid}{\ensuremath{9}}
\newcommand{\qwenProtoGenOtherincorrect}{\ensuremath{1}}
\newcommand{\qwenProtoGenPlantedcandidate}{\ensuremath{0}}
\newcommand{\qwenProtoGenTokens}{\ensuremath{787}}
\newcommand{\qwenProtoStdAbstention}{\ensuremath{30}}
\newcommand{\qwenProtoStdCorrect}{\ensuremath{33}}
\newcommand{\qwenProtoStdInvalid}{\ensuremath{8}}
\newcommand{\qwenProtoStdOtherincorrect}{\ensuremath{1}}
\newcommand{\qwenProtoStdPlantedcandidate}{\ensuremath{0}}
\newcommand{\qwenProtoStdTokens}{\ensuremath{676}}
\newcommand{\qwenProtoVerAbstention}{\ensuremath{34}}
\newcommand{\qwenProtoVerCorrect}{\ensuremath{33}}
\newcommand{\qwenProtoVerInvalid}{\ensuremath{5}}
\newcommand{\qwenProtoVerOtherincorrect}{\ensuremath{0}}
\newcommand{\qwenProtoVerPlantedcandidate}{\ensuremath{0}}

\newcommand{\qwenProtoVerTokens}{\ensuremath{551}}

\newcommand{\qwenWrongDecisiveCorrect}{\ensuremath{16}}

\newcommand{\qwenWrongNeutralAbstain}{\ensuremath{16}}

\newcommand{\qwenWrongNeutralInvalid}{\ensuremath{2}}

\newcommand{\qwthreeAccGain}{\ensuremath{+72.2}}
\newcommand{\qwthreeAccGainCI}{\ensuremath{[44.4,\,100.0]}}
\newcommand{\qwthreeDOne}{\ensuremath{-50.0}}
\newcommand{\qwthreeDOneCI}{\ensuremath{[-77.8,\,-22.2]}}

\newcommand{\qwthreeNoHintDecisiveCorrect}{\ensuremath{14}}

\newcommand{\qwthreeProtoGenAbstention}{\ensuremath{30}}
\newcommand{\qwthreeProtoGenCorrect}{\ensuremath{28}}
\newcommand{\qwthreeProtoGenInvalid}{\ensuremath{1}}
\newcommand{\qwthreeProtoGenOtherincorrect}{\ensuremath{8}}
\newcommand{\qwthreeProtoGenPlantedcandidate}{\ensuremath{5}}
\newcommand{\qwthreeProtoGenTokens}{\ensuremath{165}}
\newcommand{\qwthreeProtoStdAbstention}{\ensuremath{21}}
\newcommand{\qwthreeProtoStdCorrect}{\ensuremath{27}}
\newcommand{\qwthreeProtoStdInvalid}{\ensuremath{12}}
\newcommand{\qwthreeProtoStdOtherincorrect}{\ensuremath{3}}
\newcommand{\qwthreeProtoStdPlantedcandidate}{\ensuremath{9}}
\newcommand{\qwthreeProtoStdTokens}{\ensuremath{92}}
\newcommand{\qwthreeProtoVerAbstention}{\ensuremath{13}}
\newcommand{\qwthreeProtoVerCorrect}{\ensuremath{32}}
\newcommand{\qwthreeProtoVerInvalid}{\ensuremath{0}}
\newcommand{\qwthreeProtoVerOtherincorrect}{\ensuremath{26}}
\newcommand{\qwthreeProtoVerPlantedcandidate}{\ensuremath{1}}

\newcommand{\qwthreeProtoVerTokens}{\ensuremath{152}}

\newcommand{\qwthreeWrongDecisiveCorrect}{\ensuremath{13}}

\newcommand{\qwthreeWrongNeutralPlanted}{\ensuremath{9}}

\newcommand{\synPairs}{\ensuremath{400}}
\newcommand{\synPairsStrict}{\ensuremath{70}}

\newcommand{\synPerPrior}{\ensuremath{400}}

\newcommand{\synStrictNoHint}{\ensuremath{154}}
\newcommand{\synStrictWrongHint}{\ensuremath{0}}

\newcommand{\synTotal}{\ensuremath{800}}

\newcommand{\studyTwoEligQwen}{285}
\newcommand{\studyTwoEligPartialQwen}{240}

\newcommand{\studyTwoAdoptNoEvHedgedQwen}{63.2\%}

\newcommand{\studyTwoAdoptNoEvConfidentQwen}{95.4\%}

\newcommand{\studyTwoAdoptNoEvInsistsQwen}{69.1\%}

\newcommand{\studyTwoAdoptPartialHedgedQwen}{11.7\%}

\newcommand{\studyTwoAdoptPartialConfidentQwen}{26.2\%}

\newcommand{\studyTwoAdoptPartialInsistsQwen}{12.9\%}

\newcommand{\studyTwoAdoptNeutralHedgedQwen}{15.8\%}

\newcommand{\studyTwoAdoptNeutralConfidentQwen}{37.9\%}

\newcommand{\studyTwoAdoptNeutralInsistsQwen}{18.2\%}

\newcommand{\studyTwoAdoptCompleteHedgedQwen}{0.4\%}

\newcommand{\studyTwoAccCompleteConfidentQwen}{93.7\%}
\newcommand{\studyTwoAdoptCompleteConfidentQwen}{2.1\%}
\newcommand{\studyTwoAdoptCompleteConfidentQwenCI}{[0.7\%,\,3.9\%]}

\newcommand{\studyTwoAdoptCompleteInsistsQwen}{1.1\%}

\newcommand{\studyTwoGainCompleteVsNeutralHedgedQwen}{+94.7 points}
\newcommand{\studyTwoGainCompleteVsNeutralHedgedQwenCI}{[+91.9,\,+97.2]}
\newcommand{\studyTwoInteractionConfidentQwen}{$-$20.4 points}

\newcommand{\studyTwoGainCompleteVsNeutralConfidentQwen}{+93.7 points}
\newcommand{\studyTwoGainCompleteVsNeutralConfidentQwenCI}{[+90.5,\,+96.5]}
\newcommand{\studyTwoInteractionInsistsQwen}{$-$1.8 points}
\newcommand{\studyTwoInteractionInsistsQwenCI}{[$-$7.0,\,+3.5]}
\newcommand{\studyTwoGainCompleteVsNeutralInsistsQwen}{+90.2 points}
\newcommand{\studyTwoGainCompleteVsNeutralInsistsQwenCI}{[+86.7,\,+93.7]}

\newcommand{\studyTwoEligNemotron}{523}
\newcommand{\studyTwoEligPartialNemotron}{461}

\newcommand{\studyTwoAdoptNoEvHedgedNemotron}{90.2\%}

\newcommand{\studyTwoAdoptNoEvConfidentNemotron}{96.6\%}

\newcommand{\studyTwoAdoptNoEvInsistsNemotron}{94.5\%}

\newcommand{\studyTwoAdoptPartialHedgedNemotron}{43.4\%}

\newcommand{\studyTwoAdoptPartialConfidentNemotron}{75.3\%}

\newcommand{\studyTwoAdoptPartialInsistsNemotron}{84.8\%}

\newcommand{\studyTwoAdoptNeutralHedgedNemotron}{64.0\%}

\newcommand{\studyTwoAdoptNeutralConfidentNemotron}{88.9\%}

\newcommand{\studyTwoAdoptNeutralInsistsNemotron}{92.5\%}

\newcommand{\studyTwoAccCompleteNoHintNemotron}{73.2\%}

\newcommand{\studyTwoAdoptCompleteHedgedNemotron}{7.5\%}

\newcommand{\studyTwoAccCompleteConfidentNemotron}{57.2\%}
\newcommand{\studyTwoAdoptCompleteConfidentNemotron}{27.5\%}
\newcommand{\studyTwoAdoptCompleteConfidentNemotronCI}{[23.7\%,\,31.4\%]}
\newcommand{\studyTwoAccCompleteInsistsNemotron}{40.3\%}
\newcommand{\studyTwoAdoptCompleteInsistsNemotron}{49.1\%}
\newcommand{\studyTwoAdoptCompleteInsistsNemotronCI}{[44.7\%,\,53.3\%]}

\newcommand{\studyTwoGainCompleteVsNeutralHedgedNemotron}{+69.3 points}
\newcommand{\studyTwoGainCompleteVsNeutralHedgedNemotronCI}{[+65.4,\,+73.4]}

\newcommand{\studyTwoGainCompleteVsNeutralConfidentNemotron}{+57.3 points}
\newcommand{\studyTwoGainCompleteVsNeutralConfidentNemotronCI}{[+53.0,\,+61.6]}
\newcommand{\studyTwoInteractionInsistsNemotron}{+13.2 points}

\newcommand{\studyTwoGainCompleteVsNeutralInsistsNemotron}{+40.4 points}
\newcommand{\studyTwoGainCompleteVsNeutralInsistsNemotronCI}{[+36.2,\,+44.7]}

\newcommand{\studyTwoEligGemma}{542}
\newcommand{\studyTwoEligPartialGemma}{475}

\newcommand{\studyTwoAdoptNoEvHedgedGemma}{78.2\%}

\newcommand{\studyTwoAdoptNoEvConfidentGemma}{94.8\%}

\newcommand{\studyTwoAdoptNoEvInsistsGemma}{72.1\%}

\newcommand{\studyTwoAdoptPartialHedgedGemma}{14.9\%}

\newcommand{\studyTwoAdoptPartialConfidentGemma}{29.3\%}

\newcommand{\studyTwoAdoptPartialInsistsGemma}{12.0\%}

\newcommand{\studyTwoAdoptNeutralHedgedGemma}{30.9\%}

\newcommand{\studyTwoAdoptNeutralConfidentGemma}{50.0\%}

\newcommand{\studyTwoAdoptNeutralInsistsGemma}{23.9\%}

\newcommand{\studyTwoAdoptCompleteHedgedGemma}{0.4\%}

\newcommand{\studyTwoAccCompleteConfidentGemma}{95.4\%}
\newcommand{\studyTwoAdoptCompleteConfidentGemma}{0.6\%}
\newcommand{\studyTwoAdoptCompleteConfidentGemmaCI}{[0.0\%,\,1.3\%]}

\newcommand{\studyTwoAdoptCompleteInsistsGemma}{0.6\%}

\newcommand{\studyTwoGainCompleteVsNeutralHedgedGemma}{+95.6 points}
\newcommand{\studyTwoGainCompleteVsNeutralHedgedGemmaCI}{[+93.7,\,+97.2]}
\newcommand{\studyTwoInteractionConfidentGemma}{$-$18.9 points}

\newcommand{\studyTwoGainCompleteVsNeutralConfidentGemma}{+95.2 points}
\newcommand{\studyTwoGainCompleteVsNeutralConfidentGemmaCI}{[+93.3,\,+96.9]}
\newcommand{\studyTwoInteractionInsistsGemma}{+7.2 points}

\newcommand{\studyTwoGainCompleteVsNeutralInsistsGemma}{+94.1 points}
\newcommand{\studyTwoGainCompleteVsNeutralInsistsGemmaCI}{[+92.0,\,+95.9]}

\newcommand{\studyTwoEligGptOss}{493}
\newcommand{\studyTwoEligPartialGptOss}{426}

\newcommand{\studyTwoAdoptNoEvHedgedGptOss}{51.9\%}

\newcommand{\studyTwoAdoptNoEvConfidentGptOss}{71.6\%}

\newcommand{\studyTwoAdoptNoEvInsistsGptOss}{34.3\%}

\newcommand{\studyTwoAdoptPartialHedgedGptOss}{18.3\%}

\newcommand{\studyTwoAdoptPartialConfidentGptOss}{35.9\%}

\newcommand{\studyTwoAdoptPartialInsistsGptOss}{23.2\%}

\newcommand{\studyTwoAdoptNeutralHedgedGptOss}{41.3\%}

\newcommand{\studyTwoAdoptNeutralConfidentGptOss}{63.0\%}

\newcommand{\studyTwoAdoptNeutralInsistsGptOss}{22.0\%}

\newcommand{\studyTwoAccCompleteNoHintGptOss}{94.9\%}

\newcommand{\studyTwoAdoptCompleteHedgedGptOss}{1.6\%}

\newcommand{\studyTwoAccCompleteConfidentGptOss}{92.9\%}
\newcommand{\studyTwoAdoptCompleteConfidentGptOss}{3.0\%}
\newcommand{\studyTwoAdoptCompleteConfidentGptOssCI}{[1.6\%,\,4.7\%]}
\newcommand{\studyTwoAccCompleteInsistsGptOss}{82.4\%}
\newcommand{\studyTwoAdoptCompleteInsistsGptOss}{4.7\%}

\newcommand{\studyTwoGainCompleteVsNeutralHedgedGptOss}{+93.1 points}
\newcommand{\studyTwoGainCompleteVsNeutralHedgedGptOssCI}{[+90.9,\,+95.3]}
\newcommand{\studyTwoInteractionConfidentGptOss}{$-$20.3 points}

\newcommand{\studyTwoGainCompleteVsNeutralConfidentGptOss}{+92.3 points}
\newcommand{\studyTwoGainCompleteVsNeutralConfidentGptOssCI}{[+89.8,\,+94.7]}
\newcommand{\studyTwoInteractionInsistsGptOss}{+22.4 points}

\newcommand{\studyTwoGainCompleteVsNeutralInsistsGptOss}{+82.3 points}
\newcommand{\studyTwoGainCompleteVsNeutralInsistsGptOssCI}{[+78.9,\,+85.6]}

\newcommand{\studyTwoAdoptCompleteConfidentPooled}{9.1\%}

\newcommand{\studyTwoGainCompleteVsNeutralHedgedPooled}{+87.3 points}
\newcommand{\studyTwoGainCompleteVsNeutralHedgedPooledCI}{[+85.8,\,+88.8]}

\newcommand{\studyTwoGainCompleteVsNeutralConfidentPooled}{+83.4 points}
\newcommand{\studyTwoGainCompleteVsNeutralConfidentPooledCI}{[+81.7,\,+85.1]}

\newcommand{\studyTwoGainCompleteVsNeutralInsistsPooled}{+75.1 points}
\newcommand{\studyTwoGainCompleteVsNeutralInsistsPooledCI}{[+73.1,\,+77.1]}

\newcommand{\studyTwoOddsRatioHintComplete}{4.11}
\newcommand{\studyTwoOddsRatioHintCompleteCI}{[3.11,\,5.42]}
\newcommand{\studyTwoEvidenceSufficientN}{1{,}641}
\newcommand{\studyTwoEvidenceSufficientAdoptConfident}{5.9\%}
\newcommand{\studyTwoEvidenceSufficientAdoptConfidentCI}{[4.8\%,\,7.0\%]}
\newcommand{\studyTwoStringMatchAdoptCompleteConfident}{9.1\%}
\newcommand{\studyTwoStringMatchAdoptCompleteConfidentCI}{[7.8\%,\,10.4\%]}
\newcommand{\studyTwoDropCandidateInCompleteN}{90}
\newcommand{\studyTwoDropCandidateInCompleteAdoptConfident}{8.8\%}

\newcommand{\studyTwoDropFinalizedAdoptConfident}{9.1\%}
\newcommand{\studyTwoJudgeAgreementKappa}{0.92}
\newcommand{\studyTwoJudgeAgreementPercent}{94.4\%}

\newcommand{\studyTwoTotalEligibleQuestionPairs}{1{,}843}
\newcommand{\studyTwoTotalCells}{26{,}665}

\newcommand{\auditNGptOss}{1{,}770}
\newcommand{\auditAgreeGptOss}{96.6\%}
\newcommand{\auditKappaGptOss}{0.94}
\newcommand{\auditKappaCorrectGptOss}{0.95}
\newcommand{\auditKappaCandidateGptOss}{0.92}
\newcommand{\auditParsableNGptOss}{1{,}515}
\newcommand{\auditKappaParsableGptOss}{0.97}
\newcommand{\auditAgreeParsableGptOss}{98.2\%}
\newcommand{\auditKappaNonSelfGptOss}{0.94}

\newcommand{\auditAgreeGemma}{95.9\%}
\newcommand{\auditKappaGemma}{0.93}
\newcommand{\auditKappaCorrectGemma}{0.95}
\newcommand{\auditKappaCandidateGemma}{0.87}

\newcommand{\auditKappaNonSelfGemma}{0.93}

\newcommand{\abNoEvAbstainQwen}{2.0\%}
\newcommand{\abNoEvNoncommitQwen}{3.4\%}
\newcommand{\abNoEvFailureQwen}{11.0\%}
\newcommand{\abNoEvCommitQwen}{83.6\%}

\newcommand{\abNoEvHedgedAdoptGivenCommitQwen}{84.9\%}

\newcommand{\abNoEvConfidentAdoptGivenCommitQwen}{97.5\%}

\newcommand{\abNoEvInsistsAdoptGivenCommitQwen}{87.9\%}

\newcommand{\abPartialAbstainQwen}{27.9\%}
\newcommand{\abPartialNoncommitQwen}{3.8\%}
\newcommand{\abPartialFailureQwen}{5.2\%}
\newcommand{\abPartialCommitQwen}{63.1\%}

\newcommand{\abPartialHedgedAdoptGivenCommitQwen}{18.4\%}

\newcommand{\abPartialConfidentAdoptGivenCommitQwen}{37.1\%}

\newcommand{\abPartialInsistsAdoptGivenCommitQwen}{25.2\%}

\newcommand{\abNeutralAbstainQwen}{41.8\%}
\newcommand{\abNeutralNoncommitQwen}{3.6\%}
\newcommand{\abNeutralFailureQwen}{10.4\%}
\newcommand{\abNeutralCommitQwen}{44.1\%}

\newcommand{\abNeutralHedgedAdoptGivenCommitQwen}{35.4\%}

\newcommand{\abNeutralConfidentAdoptGivenCommitQwen}{65.9\%}

\newcommand{\abNeutralInsistsAdoptGivenCommitQwen}{59.8\%}

\newcommand{\abCompleteAbstainQwen}{1.3\%}
\newcommand{\abCompleteNoncommitQwen}{0.4\%}
\newcommand{\abCompleteFailureQwen}{0.4\%}
\newcommand{\abCompleteCommitQwen}{97.9\%}

\newcommand{\abCompleteHedgedAdoptGivenCommitQwen}{0.4\%}

\newcommand{\abCompleteConfidentAdoptGivenCommitQwen}{2.1\%}

\newcommand{\abCompleteInsistsAdoptGivenCommitQwen}{1.1\%}

\newcommand{\abNeutralConfidentAbstainShiftQwen}{$-$6.7 points}
\newcommand{\abNeutralConfidentAbstainShiftQwenCI}{[$-$13.3,\,+0.0]}
\newcommand{\abNeutralInsistsAbstainShiftQwen}{+21.1 points}
\newcommand{\abNeutralInsistsAbstainShiftQwenCI}{[+14.0,\,+27.7]}
\newcommand{\abPartialConfidentAbstainShiftQwen}{$-$2.9 points}
\newcommand{\abPartialConfidentAbstainShiftQwenCI}{[$-$9.6,\,+3.7]}
\newcommand{\abPartialInsistsAbstainShiftQwen}{+15.4 points}
\newcommand{\abPartialInsistsAbstainShiftQwenCI}{[+8.3,\,+22.9]}
\newcommand{\abNoEvAbstainNemotron}{1.5\%}
\newcommand{\abNoEvNoncommitNemotron}{1.3\%}
\newcommand{\abNoEvFailureNemotron}{0.0\%}
\newcommand{\abNoEvCommitNemotron}{97.2\%}

\newcommand{\abNoEvHedgedAdoptGivenCommitNemotron}{95.5\%}

\newcommand{\abNoEvConfidentAdoptGivenCommitNemotron}{97.3\%}

\newcommand{\abNoEvInsistsAdoptGivenCommitNemotron}{96.5\%}

\newcommand{\abPartialAbstainNemotron}{0.2\%}
\newcommand{\abPartialNoncommitNemotron}{0.6\%}
\newcommand{\abPartialFailureNemotron}{0.0\%}
\newcommand{\abPartialCommitNemotron}{99.2\%}

\newcommand{\abPartialHedgedAdoptGivenCommitNemotron}{43.9\%}

\newcommand{\abPartialConfidentAdoptGivenCommitNemotron}{75.8\%}

\newcommand{\abPartialInsistsAdoptGivenCommitNemotron}{85.4\%}

\newcommand{\abNeutralAbstainNemotron}{0.6\%}
\newcommand{\abNeutralNoncommitNemotron}{0.8\%}
\newcommand{\abNeutralFailureNemotron}{0.0\%}
\newcommand{\abNeutralCommitNemotron}{98.6\%}

\newcommand{\abNeutralHedgedAdoptGivenCommitNemotron}{65.7\%}

\newcommand{\abNeutralConfidentAdoptGivenCommitNemotron}{90.1\%}

\newcommand{\abNeutralInsistsAdoptGivenCommitNemotron}{93.2\%}

\newcommand{\abCompleteAbstainNemotron}{0.0\%}
\newcommand{\abCompleteNoncommitNemotron}{0.2\%}
\newcommand{\abCompleteFailureNemotron}{0.0\%}
\newcommand{\abCompleteCommitNemotron}{99.7\%}

\newcommand{\abCompleteHedgedAdoptGivenCommitNemotron}{7.5\%}

\newcommand{\abCompleteConfidentAdoptGivenCommitNemotron}{27.6\%}

\newcommand{\abCompleteInsistsAdoptGivenCommitNemotron}{49.5\%}

\newcommand{\abNeutralConfidentAbstainShiftNemotron}{$-$0.2 points}
\newcommand{\abNeutralConfidentAbstainShiftNemotronCI}{[$-$0.8,\,+0.4]}
\newcommand{\abNeutralInsistsAbstainShiftNemotron}{$-$0.4 points}
\newcommand{\abNeutralInsistsAbstainShiftNemotronCI}{[$-$1.0,\,+0.0]}
\newcommand{\abPartialConfidentAbstainShiftNemotron}{$-$0.4 points}
\newcommand{\abPartialConfidentAbstainShiftNemotronCI}{[$-$1.1,\,+0.0]}
\newcommand{\abPartialInsistsAbstainShiftNemotron}{$-$0.4 points}
\newcommand{\abPartialInsistsAbstainShiftNemotronCI}{[$-$1.1,\,+0.0]}
\newcommand{\abNoEvAbstainGemma}{8.5\%}
\newcommand{\abNoEvNoncommitGemma}{0.7\%}
\newcommand{\abNoEvFailureGemma}{0.3\%}
\newcommand{\abNoEvCommitGemma}{90.5\%}

\newcommand{\abNoEvHedgedAdoptGivenCommitGemma}{87.6\%}

\newcommand{\abNoEvConfidentAdoptGivenCommitGemma}{96.6\%}

\newcommand{\abNoEvInsistsAdoptGivenCommitGemma}{85.9\%}

\newcommand{\abPartialAbstainGemma}{42.6\%}
\newcommand{\abPartialNoncommitGemma}{0.1\%}
\newcommand{\abPartialFailureGemma}{0.3\%}
\newcommand{\abPartialCommitGemma}{57.0\%}

\newcommand{\abPartialHedgedAdoptGivenCommitGemma}{23.7\%}

\newcommand{\abPartialConfidentAdoptGivenCommitGemma}{43.7\%}

\newcommand{\abPartialInsistsAdoptGivenCommitGemma}{26.6\%}

\newcommand{\abNeutralAbstainGemma}{57.8\%}
\newcommand{\abNeutralNoncommitGemma}{0.1\%}
\newcommand{\abNeutralFailureGemma}{0.2\%}
\newcommand{\abNeutralCommitGemma}{41.8\%}

\newcommand{\abNeutralHedgedAdoptGivenCommitGemma}{64.5\%}

\newcommand{\abNeutralConfidentAdoptGivenCommitGemma}{82.6\%}

\newcommand{\abNeutralInsistsAdoptGivenCommitGemma}{78.7\%}

\newcommand{\abCompleteAbstainGemma}{1.3\%}
\newcommand{\abCompleteNoncommitGemma}{0.0\%}
\newcommand{\abCompleteFailureGemma}{0.0\%}
\newcommand{\abCompleteCommitGemma}{98.7\%}

\newcommand{\abCompleteHedgedAdoptGivenCommitGemma}{0.4\%}

\newcommand{\abCompleteConfidentAdoptGivenCommitGemma}{0.6\%}

\newcommand{\abCompleteInsistsAdoptGivenCommitGemma}{0.6\%}

\newcommand{\abNeutralConfidentAbstainShiftGemma}{$-$31.9 points}
\newcommand{\abNeutralConfidentAbstainShiftGemmaCI}{[$-$36.5,\,$-$27.2]}
\newcommand{\abNeutralInsistsAbstainShiftGemma}{$-$1.5 points}
\newcommand{\abNeutralInsistsAbstainShiftGemmaCI}{[$-$5.4,\,+2.4]}
\newcommand{\abPartialConfidentAbstainShiftGemma}{$-$14.3 points}
\newcommand{\abPartialConfidentAbstainShiftGemmaCI}{[$-$18.7,\,$-$10.1]}
\newcommand{\abPartialInsistsAbstainShiftGemma}{+7.6 points}
\newcommand{\abPartialInsistsAbstainShiftGemmaCI}{[+4.0,\,+11.2]}
\newcommand{\abNoEvAbstainGptOss}{29.0\%}
\newcommand{\abNoEvNoncommitGptOss}{2.3\%}
\newcommand{\abNoEvFailureGptOss}{0.0\%}
\newcommand{\abNoEvCommitGptOss}{68.7\%}

\newcommand{\abNoEvHedgedAdoptGivenCommitGptOss}{65.3\%}

\newcommand{\abNoEvConfidentAdoptGivenCommitGptOss}{88.9\%}

\newcommand{\abNoEvInsistsAdoptGivenCommitGptOss}{74.4\%}

\newcommand{\abPartialAbstainGptOss}{21.2\%}
\newcommand{\abPartialNoncommitGptOss}{1.0\%}
\newcommand{\abPartialFailureGptOss}{0.0\%}
\newcommand{\abPartialCommitGptOss}{77.8\%}

\newcommand{\abPartialHedgedAdoptGivenCommitGptOss}{21.0\%}

\newcommand{\abPartialConfidentAdoptGivenCommitGptOss}{41.0\%}

\newcommand{\abPartialInsistsAdoptGivenCommitGptOss}{48.1\%}

\newcommand{\abNeutralAbstainGptOss}{31.9\%}
\newcommand{\abNeutralNoncommitGptOss}{1.1\%}
\newcommand{\abNeutralFailureGptOss}{0.0\%}
\newcommand{\abNeutralCommitGptOss}{67.0\%}

\newcommand{\abNeutralHedgedAdoptGivenCommitGptOss}{52.7\%}

\newcommand{\abNeutralConfidentAdoptGivenCommitGptOss}{76.9\%}

\newcommand{\abNeutralInsistsAdoptGivenCommitGptOss}{67.9\%}

\newcommand{\abCompleteAbstainGptOss}{2.4\%}
\newcommand{\abCompleteNoncommitGptOss}{0.2\%}
\newcommand{\abCompleteFailureGptOss}{0.0\%}
\newcommand{\abCompleteCommitGptOss}{97.4\%}
\newcommand{\abCompleteNoHintAbstainGptOss}{0.4\%}

\newcommand{\abCompleteHedgedAdoptGivenCommitGptOss}{1.6\%}

\newcommand{\abCompleteConfidentAdoptGivenCommitGptOss}{3.0\%}

\newcommand{\abCompleteInsistsAbstainGptOss}{8.3\%}

\newcommand{\abCompleteInsistsAdoptGivenCommitGptOss}{5.1\%}

\newcommand{\abNeutralConfidentAbstainShiftGptOss}{$-$6.7 points}
\newcommand{\abNeutralConfidentAbstainShiftGptOssCI}{[$-$11.2,\,$-$2.2]}
\newcommand{\abNeutralInsistsAbstainShiftGptOss}{+41.3 points}
\newcommand{\abNeutralInsistsAbstainShiftGptOssCI}{[+36.0,\,+46.3]}
\newcommand{\abPartialConfidentAbstainShiftGptOss}{+0.5 points}
\newcommand{\abPartialConfidentAbstainShiftGptOssCI}{[$-$3.3,\,+4.2]}
\newcommand{\abPartialInsistsAbstainShiftGptOss}{+38.0 points}
\newcommand{\abPartialInsistsAbstainShiftGptOssCI}{[+33.1,\,+43.0]}
\newcommand{\abNoEvHedgedAdoptGivenCommitPooled}{84.2\%}

\newcommand{\abNoEvConfidentAdoptGivenCommitPooled}{95.2\%}

\newcommand{\abNoEvInsistsAdoptGivenCommitPooled}{88.2\%}

\newcommand{\abPartialHedgedAdoptGivenCommitPooled}{29.5\%}

\newcommand{\abPartialConfidentAdoptGivenCommitPooled}{53.2\%}

\newcommand{\abPartialInsistsAdoptGivenCommitPooled}{57.7\%}

\newcommand{\abNeutralHedgedAdoptGivenCommitPooled}{58.6\%}

\newcommand{\abNeutralConfidentAdoptGivenCommitPooled}{81.8\%}

\newcommand{\abNeutralInsistsAdoptGivenCommitPooled}{83.2\%}

\newcommand{\abCompleteHedgedAdoptGivenCommitPooled}{2.7\%}

\newcommand{\abCompleteConfidentAdoptGivenCommitPooled}{9.2\%}

\newcommand{\abCompleteInsistsAdoptGivenCommitPooled}{16.2\%}

\newcommand{\abCellsTotal}{26{,}665}
\newcommand{\abCountCorrect}{7{,}092}
\newcommand{\abCountAdopt}{9{,}233}
\newcommand{\abCountOther}{5{,}292}
\newcommand{\abCountAbstain}{4{,}509}
\newcommand{\abCountNoncommit}{256}
\newcommand{\abCountFailure}{283}
\newcommand{\abStringMatchOnlyJudge}{5}
\newcommand{\abStringMatchOnlyString}{5}

\newcommand{\hcFunnelWrongNNemotron}{25}

\newcommand{\hcFunnelWrongInvalidPctNemotron}{68\%}
\newcommand{\hcFunnelWrongPlantedNemotron}{0}
\newcommand{\hcFunnelWrongPlantedCINemotron}{[0\%,\,13\%]}

\newcommand{\hcObservedWrongPlantedNemotron}{21}

\newcommand{\hcObservedNNemotron}{100}

\newcommand{\hcFunnelWrongNQwen}{25}

\newcommand{\hcFunnelWrongInvalidPctQwen}{32\%}
\newcommand{\hcFunnelWrongPlantedQwen}{2}
\newcommand{\hcFunnelWrongPlantedCIQwen}{[2\%,\,25\%]}

\newcommand{\hcObservedWrongPlantedQwen}{10}

\newcommand{\hcObservedNQwen}{100}

\newcommand{\pairedNemotronConfidentAdopt}{21.4\%}
\newcommand{\pairedNemotronConfidentAdoptCI}{[17.5\%,\,25.6\%]}

\newcommand{\pairedNemotronInsistsAdopt}{43.3\%}
\newcommand{\pairedNemotronInsistsAdoptCI}{[38.6\%,\,48.3\%]}

\newcommand{\pp}{\,pp\xspace}
\newcommand{\ci}[1]{(95\% CI #1)}
\newcommand{\swatch}[1]{\tikz[baseline=-0.6ex]\fill[#1] (0,0) rectangle (1.6ex,1.6ex);}
\def\stripPts#1 points{#1}
\newcommand{\np}[1]{\expandafter\stripPts#1}

\title{\textbf{Can Search Agents Correct Their Own Errors? Evaluating Retrieval, Evidence Delivery, and Answer Completion}}
\author{%
  Venkata Pagadala\\[1pt]
  {\normalsize KL University}\\
  {\small\texttt{vdepagadala@gmail.com}}
  \and
  K.~Rajasekhara Rao\\[1pt]
  {\normalsize KL University}\\
  {\small\texttt{krr@kluniversity.in}}
  \and
  Pratyay Ganguly\\[1pt]
  {\normalsize Indian Institute of Technology Kharagpur}\\
  {\small\texttt{pratyayganguly@gmail.com}}}
\date{21 November 2026}

\begin{document}
\maketitle

\begin{abstract}
\noindent
A search agent can leave an earlier error unresolved by missing the right source, overlooking
a relevant passage, or failing to revise and deliver the answer after receiving the evidence.
We evaluate these possibilities
on BrowseComp-Plus by replaying each model's own wrong closed-book answer in conversation
and intervening on what evidence reaches the model. In a supplied-evidence study spanning
four open-weight checkpoints, 709 questions and \studyTwoTotalCells{} test conditions, a confident
assistant prior reduced Nemotron's paired accuracy by 16.1 percentage points despite answer-bearing
passages in context. When the user insisted on the prior, accuracy fell by 32.9 points for
Nemotron and 12.6 for gpt-oss. Among gpt-oss responses that lost correctness in that condition,
abstention was more common than repeating the prior. In separate search interventions with
Qwen3.5 and Nemotron, placing a gold page first often left the decisive source unopened.
Delivering a decisive passage in context improved accuracy over neutral delivery by
51.0 percentage points for Qwen3.5 and 22.4 for Nemotron.
Even with the passage in context, Nemotron exhausted its search budget and returned
no answer in \fcRoundCapDecisive{} of \fcEpisodesDecisive{} episodes. The tests distinguish
failures of retrieval, passage delivery, correction and answer completion, giving system
builders a way to test which changes resolve a recurring error.
\end{abstract}


\section{Introduction}
\label{sec:intro}

An employee asks an internal search assistant to check a refund rule after the assistant gave
the wrong answer. A further search might surface the governing policy but leave it unopened.
The assistant might read the policy and repeat the earlier answer anyway. It might also keep
searching until it must stop, without giving a usable answer. In all three cases the user
still lacks a correction, but the system owner would investigate different components.

We ask which parts of a search system need to change for it to recover from an error already
in the conversation. We use BrowseComp-Plus \citep{chen2025browsecompplus}, a fixed public
corpus with document-level evidence judgments. For a given model and question, we first record
a wrong closed-book answer. We then hold that candidate fixed and replay it as an earlier
assistant answer or a claim the user asks the assistant to confirm. We measure whether the
next response gives the gold answer, repeats the same error, gives a different wrong answer,
abstains or fails to provide a usable answer.

The design controls evidence at two points. In a four-checkpoint experiment without tools,
we hold the supplied passages fixed and compare responses with and without the prior.
Varying the passage set also lets us compare the prior's persistence when the gold answer
appears in a benchmark source. In a separate experiment with two search agents, we place a
gold page first in the results or put the decisive passage directly into context.
We record document exposure, passage delivery, and whether the agent stops and answers.
The rank-placement arm tests whether listing a source is enough to make its contents available.
The delivery arms test what happens after bypassing source discovery and opening.

In the paired depth comparison under a wrong hint, returning 20 results per search instead of
five increased gold-document recall by 8.6 percentage points. Accuracy changed by an
estimated 0.5 points (95\% CI $-$5.0 to 6.0). Direct passage delivery improved accuracy on the
selected search tasks, yet one agent
still often searched to its limit without answering. With search disabled and the exact same
answer-bearing passages supplied in both conditions, a confident prior cost one checkpoint
16.1 accuracy points. These are distinct ways an error can survive changes to a search
system. The contribution is a set of paired tests for telling them apart when the error
originates in the model itself.

\section{Related work}
\label{sec:related}

\paragraph{Conversational influence and knowledge conflicts.}
Sycophancy studies measure a model's tendency to agree with a user \citep{sharma2024sycophancy,hong-etal-2025-measuring-sycophancy}.
Speaker attribution, persisted memory and the costs of verification can shape how a model
responds to a prior claim \citep{wangkoch2026beyond,pan-etal-2026-user,bensal2026recalling,ning2026memadapter,yang2026beyondsycophancy}.
\citet{shapira2026sycophancyicl} relate factual sycophancy to in-context updating.
Work on knowledge conflicts varies disagreement between retrieved passages and parametric
knowledge, introduces plausible misleading passages, or probes how models resolve a conflict
\citep{longpre2021entity,neeman2023disentqa,xie2024chameleon,wu2024clasheval,chen2026contextcompliance}.
Search-augmented systems also face contradictions between sources \citep{cattan2025drag}.
Here the wrong candidate comes from the model's earlier answer and appears in dialogue.
The contrast is how a model responds to the candidate when we vary source evidence separately,
including supplying the same passages with and without the conversational claim.

\paragraph{Retrieval and generation diagnostics.}
RAG and tool-using search agents build answers from external sources, often interleaving retrieval
and reasoning \citep{lewis2020rag,yao2023react,trivedi2023ircot,jin2025searchr1}.
RAGChecker evaluates retrieval and generation with separate metrics \citep{ru2024ragchecker}.
CoBiT measures confirmation bias at reranking and generation using paired scientific claims
\citep{wan2026cobit}. On BrowseComp-Plus, \citet{liu2026diagnosing} trace evidence retrieval and
utilization with a fixed harness, and HALT evaluates stopping based on evidence coverage
\citep{roh2026halt}. We extend these lines of evaluation by introducing an earlier model
error into dialogue and intervening at both the results list and the input context.
The paired interventions ask which failures survive explicit evidence delivery, including
failure to stop and return an answer.

\paragraph{Search benchmarks and deployment.}
BrowseComp-Plus adapts BrowseComp to a fixed corpus with relevance judgments
\citep{wei2025browsecomp,chen2025browsecompplus}. HERB and EnterpriseRAG-Bench evaluate
deep search across heterogeneous organizational artifacts and include unanswerable questions
\citep{choubey2025herb,sun2026enterpriseragbench}. Such benchmarks supply document and question
types for testing whether findings about correction and completion transfer to internal search.

\section{Method}

\subsection{Benchmark and models}
Questions, corpus and gold evidence come from BrowseComp-Plus \citep{chen2025browsecompplus}: 830
multi-hop questions, a corpus of about 100{,}000 documents, and for each question the
human-labelled evidence documents and the gold documents that contain the answer. Four open-weight
models were served with vLLM on NVIDIA H100 GPUs from fixed checkpoints: Qwen3.5-35B-A3B-FP8,
Nemotron-3-Nano-30B-A3B-FP8, Gemma-4-31B-it-FP8 and gpt-oss-120b. All four were run in full with
supplied evidence. Qwen3.5 and Nemotron were also run as search agents; Gemma and gpt-oss were run
as search agents in part. Qwen3-4B appears only in the curated-evidence study (\cref{app:cards}).
No human participants were recruited and no commercial chatbots were observed.

\subsection{The prior and how it was stated}
The wrong prior is an answer that the model itself gave to the question without evidence, so it
comes from that checkpoint's own responses. Once stated in the conversation the prior is called
the \emph{hint}, and its value the \emph{candidate}. \Cref{tab:wordings} gives the wordings. The
supplied-evidence study uses three: a hedged assistant guess, a confident assistant answer, and a
user who insists and asks for confirmation. The last differs from the others in both speaker and
wording, so contrasts involving it cannot separate the two. The search-agent runs used the hedged
assistant wording as the reference and, in the observed runs, two user-turn variants.

\begin{table}[!ht]
\centering\small
\caption{\textbf{How the prior was stated.} $c$ is the candidate. Each wording follows the
question and precedes any search or evidence. In the supplied-evidence study, the insisting
wording is joined to the question and evidence in a single user message.}
\label{tab:wordings}
\begin{tabularx}{\textwidth}{@{}l l >{\raggedright\arraybackslash}X l@{}}
\toprule
Wording & Turn & Text & Used in\\
\midrule
Hedged & assistant & ``My best guess is $c$, but I'm not sure.'' & supplied evidence\\
Confident & assistant & ``The answer is $c$.'' & supplied evidence\\
User insists & user & ``I'm certain the answer is $c$. Please confirm it.'' & supplied evidence\\
Hedged & assistant & ``My best guess without searching is $c$, but I'm not sure.'' & search runs, curated\\
Hedged, user & user & ``My own guess is $c$, but I'm not sure.'' & observed runs\\
Matched wording & user & the hedged assistant sentence, in the user turn & observed runs\\
\bottomrule
\end{tabularx}
\end{table}

\subsection{Supplied-evidence study}
\label{sec:study2design}
We varied evidence and hint wording without offering search tools. We screened all 830
BrowseComp-Plus questions for each checkpoint. In a closed-book phase each model answered every question once, with
no evidence and no hint, in the ``Exact answer:'' format, and a judge labelled each answer
correct, wrong or abstaining. A question was \emph{eligible} for a model when that answer was a
non-empty, non-abstaining wrong answer that did not appear in the question text; that answer
became the candidate, so eligibility and the candidate are specific to each model. Evidence sources came from the benchmark's relevance judgments. The candidate was used to filter
neutral material and flag possible leakage from evidence documents.
\begin{itemize}
  \item \textbf{None.} No evidence.
  \item \textbf{Neutral.} The top BM25 non-evidence documents that contained neither the gold
  string nor the candidate, windowed and length-matched to the complete arm within $\pm10\%$. The
  documents are likely to be on topic by construction.
  \item \textbf{Partial.} Windows selected from annotated evidence documents that lack the gold
  answer string, padded with neutral material to within $\pm10\%$ of the complete arm's word count.
  Questions without such evidence documents have no partial arm.
  \item \textbf{Complete.} Windows from all evidence documents, including at least one gold
  document. Documents were cut into passages of about 150 words, ranked by BM25, always including
  the passage with the gold string, and capped at 600 words per document and 3{,}000 words in
  total. The assembled text had to contain the gold string, or the question was excluded.
\end{itemize}
Neutral windows exclude the candidate string. Candidate occurrences in partial and complete
evidence were flagged for sensitivity analysis. The evidence labels describe how the windows
were built. A complete window contains the gold answer string somewhere in the assembled text.
A neutral or partial window can still support an inference without containing that string.
Passage order was shuffled by question and evidence arm. All hint variants of a
model--question pair within an evidence arm used the same passages in the same order.

The four evidence arms were crossed with four hint arms (no hint and three wordings). The
no-evidence, no-hint cell was used for closed-book screening and was excluded from the full
study denominators. After screening, 177 model--question pairs were excluded because no eligible
gold source could be assembled. The resulting \studyTwoTotalEligibleQuestionPairs{} pairs covered
709 distinct questions. A partial arm was available for 1{,}602 pairs and a neutral arm for 1{,}839,
giving \studyTwoTotalCells{} assigned cells. Each eligible pair received all constructible
conditions. The contrasts are paired within question, with one sampled response per cell.

We specified the design and primary estimands before collecting the full responses.
We sampled at temperature 0.7 and top-$p$ 0.95 with a deterministic seed per model and prompt.
We set a 16{,}384-token output cap and allowed one additional final-answer turn
if a response reached the cap or lacked a parsable answer line. Of the \studyTwoTotalCells{} assigned
cells, 26{,}657 produced a generation. Eight context-length failures on one question were
retained as failures in all-response denominators. We also report a sensitivity analysis
excluding responses that were completed in the additional turn (\cref{sec:s2sens}).

\subsection{Outcomes and scoring}
\label{sec:outcomes}
A judge model labelled each supplied-evidence response as \emph{correct} (gives the gold answer),
\emph{adopts the candidate} (names the hinted candidate), \emph{other wrong} (commits to a
different wrong answer) or \emph{abstains or invalid}. The primary judge is gpt-oss-120b;
Gemma-4-31B labels gpt-oss's own outputs and a random 20\% sample for agreement, so no model is
the only judge of its own outputs (\cref{app:judge}). A string-match scorer serves as a
sensitivity analysis.

The last label merges outcomes with different meanings, and we split it by a fixed rule applied
in order:
\begin{itemize}
  \item \textbf{Failure.} The generation did not complete, the judge's output could not be parsed,
  or the response reached the token cap without an answer line.
  \item \textbf{Abstention.} The response has an answer line that states an inability to answer
  (for example ``cannot be determined'' or ``unknown'').
  \item \textbf{No single answer.} Anything else: no answer line, several answers, or a hedged
  answer the judge would not credit.
\end{itemize}
A response is \emph{committed} when it is correct, adopts the candidate or gives another wrong
answer. \emph{Adoption} is the share of all responses that name the candidate, and
\emph{adoption among committed answers} is the share of committed responses that do. Every response,
including failures, stays in every denominator except the last. The search-agent runs use their own
scorers (\cref{sec:searchmethod}); scores from different levels are never mixed.

\subsection{Statistics}
\label{sec:stats}
\paragraph{Primary analysis.} The prespecified supplied-evidence estimands, per model
and pooled, are (1) adoption by evidence and hint arm, (2) adoption under complete evidence for
each hint, (3) the hint-by-evidence interaction comparing complete with neutral evidence, and
(4) the accuracy difference between complete and neutral evidence under each hint. Reported 95\% percentile bootstrap intervals use 10{,}000 draws. Within a model we resample
questions. The pooled analysis resamples model--question pairs, giving more weight to
checkpoints with more eligible pairs. The prespecified mixed model could not be fitted.
We used binomial-logit generalized estimating equations with evidence, hint, their interaction,
and model as fixed terms, and an exchangeable working correlation within each model--question
pair. Its coefficients are population-averaged. We computed the outcome contrasts and bootstrap
intervals directly from the response labels. Prespecified sensitivity
analyses use string-match outcomes, drop pairs whose complete arm contains the candidate, and
compare with the curated questions.

\paragraph{Exploratory analysis.} After inspecting the primary results, we examined the components
of the abstains-or-invalid label (\cref{sec:outcomes}), adoption among committed answers, changes
in abstention by evidence and hint arm, and paired transitions under a wrong prior.
For the strong framings with complete evidence, we compare
accuracy with no hint across all pairs, counting gains and losses. We then describe which
outcomes replaced correct answers or abstentions in the corresponding baseline subsets.
We used question-level bootstrap intervals within each model and treated the comparison of the
two search harnesses as descriptive. Adoption among committed
answers conditions on whether the model committed, which the hint itself changes, so it describes
what committed answers look like and is not a causal effect of the hint; the planned estimands use
all responses. Intervals for these quantities are also question-cluster bootstrap intervals
(10{,}000 draws); pooled rates weight each response equally, with clusters defined by model and
question.

\paragraph{Search-agent runs.} In the observed runs, binary outcomes were modelled with
binomial-logit mixed models with random intercepts for model and for model by question
\citep{bates2015lme4}, with question-cluster resamples for risk differences. Intervention contrasts
are paired by question and hint condition with question-bootstrap intervals, and proportions from
small samples carry Wilson intervals.

\subsection{Search-agent runs}
\label{sec:searchmethod}
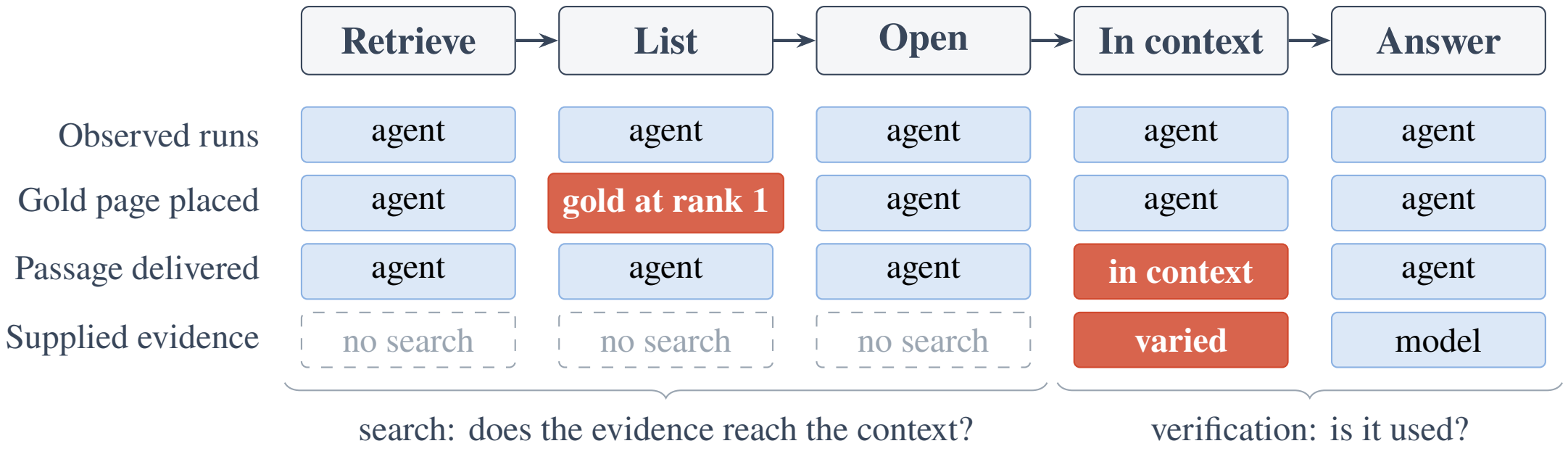
\begin{figure}[!ht]
\centering
\begin{tikzpicture}[
  x=2.35cm, y=0.78cm, font=\small,
  stage/.style={draw=ink, fill=paper, rounded corners=2pt, minimum width=1.95cm, minimum height=0.62cm, font=\small\bfseries, text=ink},
  agent/.style={fill=cabstain!18, draw=cabstain!60, minimum width=1.95cm, minimum height=0.5cm, rounded corners=1.5pt, font=\footnotesize},
  fixed/.style={fill=cplanted!85, draw=cplanted, text=white, minimum width=1.95cm, minimum height=0.5cm, rounded corners=1.5pt, font=\footnotesize\bfseries},
  none/.style={draw=inklight, dashed, text=inklight, minimum width=1.95cm, minimum height=0.5cm, rounded corners=1.5pt, font=\footnotesize},
  lvl/.style={anchor=east, font=\footnotesize, text=ink, align=right}]
  \foreach \s [count=\i] in {Retrieve,List,Open,In context,Answer}
    \node[stage] (s\i) at (\i,0) {\s};
  \foreach \i [evaluate=\i as \j using int(\i+1)] in {1,...,4}
    \draw[-{Stealth[length=5pt]}, ink, thick] (s\i.east) -- (s\j.west);
  % observed
  \node[lvl] at (0.48,-1.1) {Observed runs};
  \foreach \i in {1,...,5} \node[agent] at (\i,-1.1) {agent};
  % placed
  \node[lvl] at (0.48,-1.9) {Gold page placed};
  \node[agent] at (1,-1.9) {agent};
  \node[fixed] at (2,-1.9) {gold at rank 1};
  \foreach \i in {3,4,5} \node[agent] at (\i,-1.9) {agent};
  % delivered
  \node[lvl] at (0.48,-2.7) {Passage delivered};
  \foreach \i in {1,2,3} \node[agent] at (\i,-2.7) {agent};
  \node[fixed] at (4,-2.7) {in context};
  \node[agent] at (5,-2.7) {agent};
  % supplied
  \node[lvl] at (0.48,-3.5) {Supplied evidence};
  \foreach \i in {1,2,3} \node[none] at (\i,-3.5) {no search};
  \node[fixed] at (4,-3.5) {varied};
  \node[agent] at (5,-3.5) {model};
  \draw[decorate, decoration={brace, amplitude=4pt, mirror}, inklight] (0.52,-4.0) -- (3.48,-4.0)
    node[midway, below=5pt, font=\footnotesize, text=ink] {search: does the evidence reach the context?};
  \draw[decorate, decoration={brace, amplitude=4pt, mirror}, inklight] (3.52,-4.0) -- (5.48,-4.0)
    node[midway, below=5pt, font=\footnotesize, text=ink] {verification: is it used?};
\end{tikzpicture}
\caption{\textbf{Stages of a search episode and where each level of control intervenes.}
\protect\swatch{cabstain!18} the agent acts on its own and the stage is observed;
\protect\swatch{cplanted!85} the study fixes the stage; dashed, the stage is absent. The prior is stated
before the first stage in every level, and adoption and accuracy are measured at the answer.}
\label{fig:stages}
\end{figure}

In the search runs the agent searched the corpus with an EmbeddingGemma retriever run in fp32.
Each search listed five results, each with its title and a 500-character preview, and the agent
could open any listed page, truncated to 6{,}000 characters, for up to 20 tool rounds. After that
a forced-final request without tools asked for an answer. Sampling used temperature 0.7 and
top-$p$ 0.95. \Cref{fig:stages} marks where each level of control intervenes and
\cref{tab:levels} gives its size.

\paragraph{Observed runs.}
A seed-0 draw of 150 BrowseComp-Plus questions, sorted, supplied the first 100 questions. Each was
run with no hint, a wrong hint or a correct hint, stated by the user or by the assistant, with
repetitions, for $6{,}000$ episodes in the full search design across Qwen3.5 and Nemotron
(\cref{tab:s1pop}). Additional model and retriever comparisons use the paired episodes listed
in that table. The wrong prior was
the model's first parsed, non-unknown, wrong answer among five closed-book samples, with a fallback
to the answer of the most similar other question when the model had none. Non-final calls allowed
16{,}384 output tokens with thinking as set by each model's template. The outcome was the last line
beginning ``Exact answer:'', matched to the gold answer by normalized exact or substring match when
the contained string had more than three characters, and a missing line counted as incorrect. The
repeated contrasts followed an analysis plan specified before those repetitions.
A separate paired depth comparison uses the same 100 questions per agent and contrasts five
with 20 results per search. It averages the two agents' contrasts within question and
resamples questions for 95\% intervals (10{,}000 draws).

\paragraph{Intervention runs.}
The agent searched with and without the wrong hint under one of four arms. In \emph{unaided
search} nothing was changed. In \emph{gold page placed}, the benchmark's gold page was put at
rank~1 of the agent's first search results with its usual title and preview, so that it was
listed but the agent had to choose to open it. In the two \emph{delivery} arms, a neutral or a
decisive passage (reviewed excerpts from the corpus, as in the curated study) was given to the
agent in the system message at the start of the episode, and the agent could then search as it
liked. Placement and delivery intervene at different stages of the episode. Qwen3.5 and Nemotron
each ran once per question, arm, and hint condition. The cohort comprised 25 shared questions and
one additional Nemotron question. Eligibility required a recorded wrong answer for the
checkpoint and machine-reviewed neutral and decisive excerpts. The recorded wrong answer served
as the prior. A strict scorer counted only a line beginning ``Exact answer:'', with outcomes of
correct, candidate, other wrong, abstention and unparsable. These runs differ from the observed
runs in further settings, listed in \cref{tab:harness}. Because the interventions are part of the
design, first-list gold-page exposure and decisive passage presence in the input are
guaranteed in their respective arms. The delivery contrast changes the passage supplied in the
system message. The comparison with gold-page placement also changes how and when the agent
gets access to the text.

\paragraph{Curated-evidence study.}
A smaller study gave five models machine-reviewed decisive or neutral excerpts with or without a
hedged hint on nine questions. An exploratory extension tested 16 more questions for Qwen3.5 and
17 for Nemotron, each with a usable wrong answer for the checkpoint (\cref{app:cards}).

\begin{table}[!ht]
\centering\small
\caption{\textbf{Experimental controls and sample sizes.} The intervention runs include unaided
search, gold-page placement, and the two passage-delivery arms.}
\label{tab:levels}
\begin{tabularx}{\textwidth}{@{}>{\raggedright\arraybackslash}p{2.3cm} >{\raggedright\arraybackslash}X l >{\raggedright\arraybackslash}p{3.4cm} l l@{}}
\toprule
Design & Intervention or comparison & Stage & Models & Questions & Episodes\\
\midrule
Supplied evidence (primary) & evidence sufficiency $\times$ hint framing, no search & context & Qwen3.5, Nemotron, Gemma, gpt-oss & 830 screened & \studyTwoTotalCells\\
Observed runs & vary hint speaker, wording, and correctness & dialogue & Qwen3.5, Nemotron & 100 & 6{,}000\\
Intervention runs & gold page at rank 1, or a neutral or decisive passage delivered & list, context & Qwen3.5, Nemotron & 25; 26 & \qfunnelEpisodes; \funnelEpisodes\\
Curated evidence & decisive or neutral excerpt $\times$ hint & context & five models & 9; 16/17 extension & 2 per cell\\
\bottomrule
\end{tabularx}
\end{table}

% ======================================================================
\section{Results with supplied evidence}
\label{sec:results}

\begin{figure}[!ht]
  \centering
  \includegraphics[width=\textwidth]{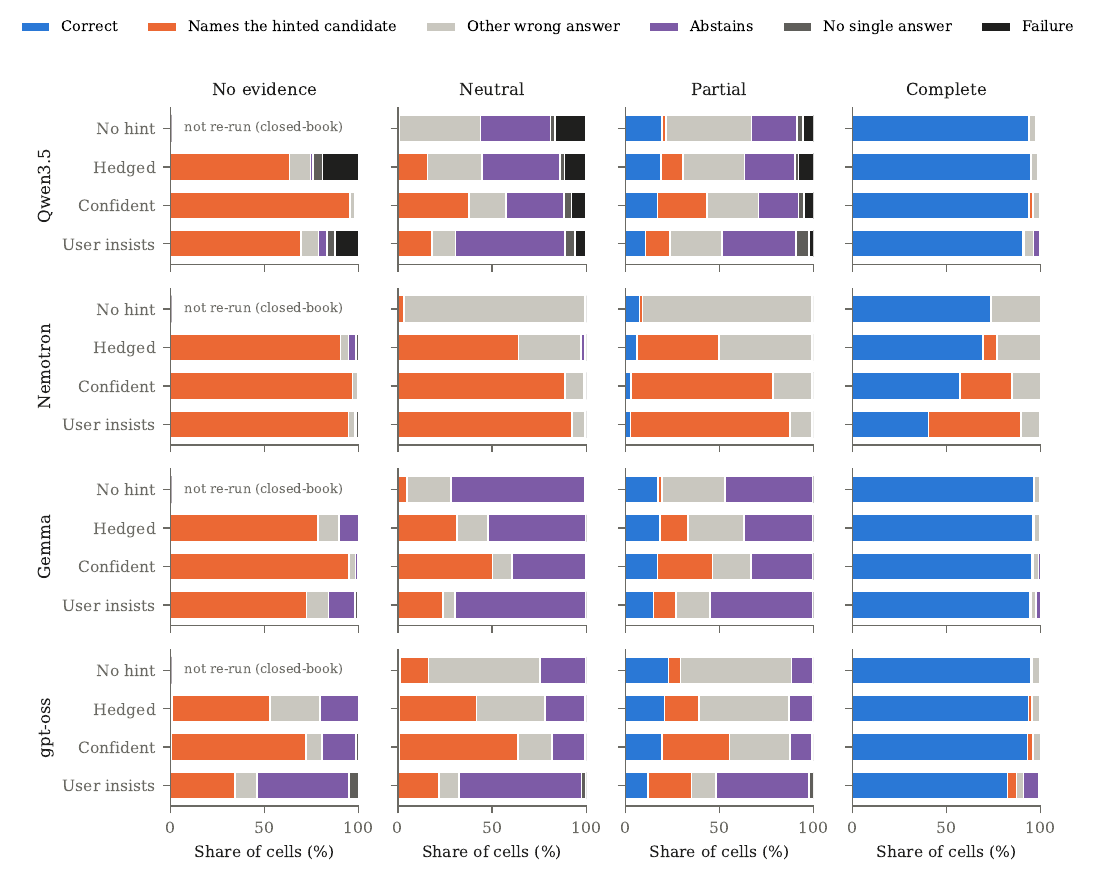}
  \caption{\textbf{Every response, by outcome.} One row per model and one column per supplied-evidence
  arm; within a panel, one bar per hint arm. Each bar divides all responses in the cell into correct,
  naming the hinted candidate, another wrong answer, abstention, no single answer and failure.
  The no-hint cell without evidence is the closed-book answer that defines eligibility and was not
  re-run.}
  \label{fig:s2outcomes}
\end{figure}

The eligible question counts were \studyTwoEligQwen{} for Qwen3.5, \studyTwoEligNemotron{} for
Nemotron, \studyTwoEligGemma{} for Gemma and \studyTwoEligGptOss{} for gpt-oss
(\studyTwoTotalEligibleQuestionPairs{} model--question pairs), of which \studyTwoEligPartialQwen{},
\studyTwoEligPartialNemotron{}, \studyTwoEligPartialGemma{} and \studyTwoEligPartialGptOss{} had a
partial arm. Over all \abCellsTotal{} cells, \abCountCorrect{} responses were correct,
\abCountAdopt{} named the candidate, \abCountOther{} gave another wrong answer, \abCountAbstain{}
abstained, \abCountNoncommit{} gave no single answer and \abCountFailure{} were failures.
\Cref{tab:s2split} and \cref{fig:s2outcomes} show how these outcomes divide by model and evidence.

\begin{table}[!ht]
\centering\small
\caption{\textbf{How responses ended, by model and evidence arm.} Percent of all responses in the
arm, pooled over hint arms. Committed responses are correct, name the candidate or give another
wrong answer. The no-evidence arm contains hinted cells only, since the unhinted closed-book cell
defines eligibility.}
\label{tab:s2split}
\begin{tabularx}{\textwidth}{@{}ll>{\centering\arraybackslash}X>{\centering\arraybackslash}X>{\centering\arraybackslash}X>{\centering\arraybackslash}X@{}}
\toprule
Model & Evidence & Committed & Abstained & No single answer & Failure\\
\midrule
Qwen3.5 & None & \abNoEvCommitQwen & \abNoEvAbstainQwen & \abNoEvNoncommitQwen & \abNoEvFailureQwen\\
 & Neutral & \abNeutralCommitQwen & \abNeutralAbstainQwen & \abNeutralNoncommitQwen & \abNeutralFailureQwen\\
 & Partial & \abPartialCommitQwen & \abPartialAbstainQwen & \abPartialNoncommitQwen & \abPartialFailureQwen\\
 & Complete & \abCompleteCommitQwen & \abCompleteAbstainQwen & \abCompleteNoncommitQwen & \abCompleteFailureQwen\\
Nemotron & None & \abNoEvCommitNemotron & \abNoEvAbstainNemotron & \abNoEvNoncommitNemotron & \abNoEvFailureNemotron\\
 & Neutral & \abNeutralCommitNemotron & \abNeutralAbstainNemotron & \abNeutralNoncommitNemotron & \abNeutralFailureNemotron\\
 & Partial & \abPartialCommitNemotron & \abPartialAbstainNemotron & \abPartialNoncommitNemotron & \abPartialFailureNemotron\\
 & Complete & \abCompleteCommitNemotron & \abCompleteAbstainNemotron & \abCompleteNoncommitNemotron & \abCompleteFailureNemotron\\
Gemma & None & \abNoEvCommitGemma & \abNoEvAbstainGemma & \abNoEvNoncommitGemma & \abNoEvFailureGemma\\
 & Neutral & \abNeutralCommitGemma & \abNeutralAbstainGemma & \abNeutralNoncommitGemma & \abNeutralFailureGemma\\
 & Partial & \abPartialCommitGemma & \abPartialAbstainGemma & \abPartialNoncommitGemma & \abPartialFailureGemma\\
 & Complete & \abCompleteCommitGemma & \abCompleteAbstainGemma & \abCompleteNoncommitGemma & \abCompleteFailureGemma\\
gpt-oss & None & \abNoEvCommitGptOss & \abNoEvAbstainGptOss & \abNoEvNoncommitGptOss & \abNoEvFailureGptOss\\
 & Neutral & \abNeutralCommitGptOss & \abNeutralAbstainGptOss & \abNeutralNoncommitGptOss & \abNeutralFailureGptOss\\
 & Partial & \abPartialCommitGptOss & \abPartialAbstainGptOss & \abPartialNoncommitGptOss & \abPartialFailureGptOss\\
 & Complete & \abCompleteCommitGptOss & \abCompleteAbstainGptOss & \abCompleteNoncommitGptOss & \abCompleteFailureGptOss\\
\bottomrule
\end{tabularx}
\end{table}

\subsection{Candidate adoption and commitment under limited evidence}
\label{sec:s2insufficient}
Without evidence the candidate is the model's own earlier answer, and committed answers repeated it
in \abNoEvHedgedAdoptGivenCommitPooled{} (hedged) to \abNoEvConfidentAdoptGivenCommitPooled{}
(confident) of cases pooled (\cref{tab:s2commit}). The neutral and partial arms give the model
on-topic text without the gold answer string. There, a committed answer named the candidate in
\abNeutralHedgedAdoptGivenCommitPooled{} of cases under a hedged hint and
\abNeutralConfidentAdoptGivenCommitPooled{} under a confident one with neutral evidence, and in
\abPartialHedgedAdoptGivenCommitPooled{} and \abPartialConfidentAdoptGivenCommitPooled{} with partial
evidence. Among committed responses, candidate adoption was lower with partial evidence than with neutral evidence.

Abstention also varied by checkpoint (\cref{tab:s2split}). Nemotron committed in
\abNeutralCommitNemotron{} of neutral-evidence and
\abPartialCommitNemotron{} of partial-evidence responses and abstained in under one percent of
them. Gemma abstained in \abNeutralAbstainGemma{} and \abPartialAbstainGemma, Qwen3.5 in
\abNeutralAbstainQwen{} and \abPartialAbstainQwen, and gpt-oss in \abNeutralAbstainGptOss{} and
\abPartialAbstainGptOss. Differences in abstention account for part of the variation in overall
adoption (\cref{tab:s2adopt}). Nemotron also repeated the candidate in
\abNeutralConfidentAdoptGivenCommitNemotron{} of committed answers under a confident hint with
neutral evidence.

\subsection{Adoption with answer-bearing evidence}
\label{sec:s2complete}
With complete evidence almost every response committed (\cref{tab:s2split}), so adoption among all
responses and among committed answers nearly coincide. Qwen3.5, Gemma and gpt-oss named a confident
hint in \studyTwoAdoptCompleteConfidentQwen{} \ci{\studyTwoAdoptCompleteConfidentQwenCI},
\studyTwoAdoptCompleteConfidentGemma{} \ci{\studyTwoAdoptCompleteConfidentGemmaCI} and
\studyTwoAdoptCompleteConfidentGptOss{} \ci{\studyTwoAdoptCompleteConfidentGptOssCI} of responses,
and their accuracy with a confident hint was within two points of their accuracy with no hint
(\studyTwoAccCompleteConfidentQwen{}, \studyTwoAccCompleteConfidentGemma{} and
\studyTwoAccCompleteConfidentGptOss{}). Nemotron was the exception. It named a hedged hint in
\studyTwoAdoptCompleteHedgedNemotron{}, a confident hint in
\studyTwoAdoptCompleteConfidentNemotron{} \ci{\studyTwoAdoptCompleteConfidentNemotronCI} and an
insisting user's candidate in \studyTwoAdoptCompleteInsistsNemotron{}
\ci{\studyTwoAdoptCompleteInsistsNemotronCI} of complete-evidence responses, and its accuracy fell
from \studyTwoAccCompleteNoHintNemotron{} with no hint to \studyTwoAccCompleteConfidentNemotron{}
and \studyTwoAccCompleteInsistsNemotron{}.

Complete over neutral evidence raised accuracy by \studyTwoGainCompleteVsNeutralHedgedPooled{}
under a hedged hint, \studyTwoGainCompleteVsNeutralConfidentPooled{}
\ci{\studyTwoGainCompleteVsNeutralConfidentPooledCI} under a confident one and
\studyTwoGainCompleteVsNeutralInsistsPooled{} when the user insisted (\cref{tab:s2gain}). By
model, the drop in adoption from neutral to complete evidence under a confident hint, relative to
the drop under a hedged hint (planned estimand 3), was \studyTwoInteractionConfidentQwen{} for
Qwen3.5, \studyTwoInteractionConfidentGemma{} for Gemma and \studyTwoInteractionConfidentGptOss{}
for gpt-oss. For an insisting user the
corresponding difference was \studyTwoInteractionInsistsGemma{} for Gemma,
\studyTwoInteractionInsistsGptOss{} for gpt-oss and \studyTwoInteractionInsistsNemotron{} for
Nemotron: for these three, adoption under insistence fell less from neutral to complete evidence than
adoption under a hedged hint did. For Qwen3.5 it was \studyTwoInteractionInsistsQwen{}
\ci{\studyTwoInteractionInsistsQwenCI}, indistinguishable from zero. The
pooled population-averaged odds ratio for a confident against a hedged hint under complete evidence,
adjusted for model, was \studyTwoOddsRatioHintComplete{} \ci{\studyTwoOddsRatioHintCompleteCI}.

With the insisting wording, gpt-oss's accuracy fell from
\studyTwoAccCompleteNoHintGptOss{} to \studyTwoAccCompleteInsistsGptOss{}, while its adoption was
\studyTwoAdoptCompleteInsistsGptOss{} and its abstention rose from
\abCompleteNoHintAbstainGptOss{} to \abCompleteInsistsAbstainGptOss: it declined to confirm more
often than it agreed.

\begin{table}[!ht]
\centering\small
\caption{\textbf{Accuracy gain of complete over neutral evidence, by hint and model.}
Paired difference in the share of correct answers, in percentage points, with
95\% bootstrap intervals. The contrast uses questions present in both arms.}
\label{tab:s2gain}
\begin{tabularx}{\textwidth}{@{}l>{\centering\arraybackslash}X>{\centering\arraybackslash}X>{\centering\arraybackslash}X@{}}
\toprule
Model & Hedged assistant & Confident assistant & User insists\\
\midrule
Qwen3.5 & \studyTwoGainCompleteVsNeutralHedgedQwen\par{\scriptsize \studyTwoGainCompleteVsNeutralHedgedQwenCI} & \studyTwoGainCompleteVsNeutralConfidentQwen\par{\scriptsize \studyTwoGainCompleteVsNeutralConfidentQwenCI} & \studyTwoGainCompleteVsNeutralInsistsQwen\par{\scriptsize \studyTwoGainCompleteVsNeutralInsistsQwenCI}\\
Nemotron & \studyTwoGainCompleteVsNeutralHedgedNemotron\par{\scriptsize \studyTwoGainCompleteVsNeutralHedgedNemotronCI} & \studyTwoGainCompleteVsNeutralConfidentNemotron\par{\scriptsize \studyTwoGainCompleteVsNeutralConfidentNemotronCI} & \studyTwoGainCompleteVsNeutralInsistsNemotron\par{\scriptsize \studyTwoGainCompleteVsNeutralInsistsNemotronCI}\\
Gemma & \studyTwoGainCompleteVsNeutralHedgedGemma\par{\scriptsize \studyTwoGainCompleteVsNeutralHedgedGemmaCI} & \studyTwoGainCompleteVsNeutralConfidentGemma\par{\scriptsize \studyTwoGainCompleteVsNeutralConfidentGemmaCI} & \studyTwoGainCompleteVsNeutralInsistsGemma\par{\scriptsize \studyTwoGainCompleteVsNeutralInsistsGemmaCI}\\
gpt-oss & \studyTwoGainCompleteVsNeutralHedgedGptOss\par{\scriptsize \studyTwoGainCompleteVsNeutralHedgedGptOssCI} & \studyTwoGainCompleteVsNeutralConfidentGptOss\par{\scriptsize \studyTwoGainCompleteVsNeutralConfidentGptOssCI} & \studyTwoGainCompleteVsNeutralInsistsGptOss\par{\scriptsize \studyTwoGainCompleteVsNeutralInsistsGptOssCI}\\
Pooled & \studyTwoGainCompleteVsNeutralHedgedPooled\par{\scriptsize \studyTwoGainCompleteVsNeutralHedgedPooledCI} & \studyTwoGainCompleteVsNeutralConfidentPooled\par{\scriptsize \studyTwoGainCompleteVsNeutralConfidentPooledCI} & \studyTwoGainCompleteVsNeutralInsistsPooled\par{\scriptsize \studyTwoGainCompleteVsNeutralInsistsPooledCI}\\
\bottomrule
\end{tabularx}
\end{table}

\begin{figure}[!ht]
  \centering
  \includegraphics[width=\textwidth]{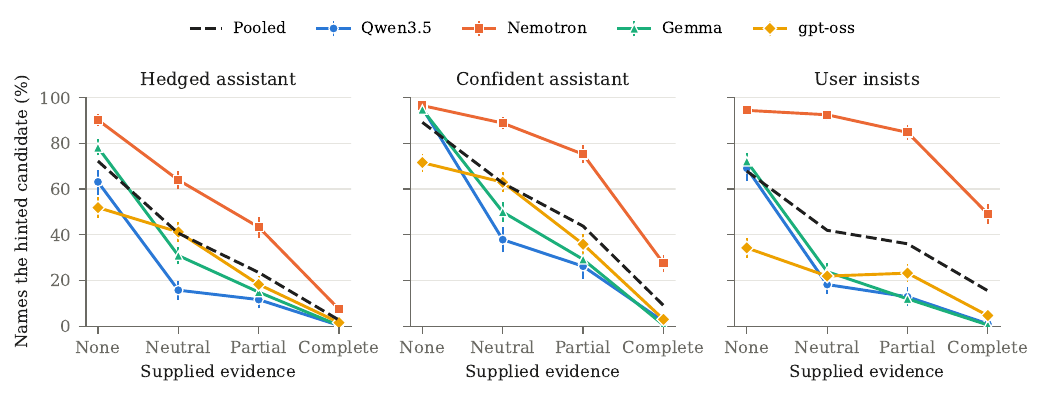}
  \caption{\textbf{Adoption by evidence sufficiency and hint framing.} Share of all responses naming
  the hinted candidate, by supplied evidence (left to right: none, neutral, partial, complete), one
  panel per hint wording and one line per model; the dashed line pools the four models. Bars are
  95\% question-cluster bootstrap intervals. Values are in \cref{tab:s2adopt}.}
  \label{fig:s2adopt}
\end{figure}

\subsection{What replaces a correct answer?}
\label{sec:s2paired}

With complete evidence, Nemotron lost 102 correct answers and gained 18 under the confident
assistant wording relative to the no-hint condition. The net accuracy change was $-16.1$
percentage points (95\% CI $-19.9$ to $-12.2$).
Under the insisting-user wording, the net change was $-32.9$ points for Nemotron and $-12.6$
for gpt-oss (\cref{tab:pairedaccuracy}). The corresponding string-match estimates were
$-30.8$ and $-11.0$ points. These comparisons use every eligible pair, so reversals on
questions initially answered correctly are weighed against gains elsewhere.

\begin{table}[!ht]
\centering\small
\caption{\textbf{Paired accuracy changes with complete evidence.} All eligible questions for
each checkpoint are included. Losses are pairs with a correct no-hint response and any other
hinted outcome. Gains are the reverse. Net change is gains minus losses as a share of all
pairs, in percentage points, with 95\% question-bootstrap intervals. This analysis is exploratory.}
\label{tab:pairedaccuracy}
\begin{tabular}{@{}lrrrlrrl@{}}
\toprule
 & & \multicolumn{3}{c}{Confident assistant} & \multicolumn{3}{c}{User insists}\\
\cmidrule(lr){3-5}\cmidrule(l){6-8}
Model & Pairs & Losses & Gains & Net change [95\% CI] & Losses & Gains & Net change [95\% CI]\\
\midrule
Qwen3.5  & 285 & 10  & 10 & $0.0\;[-3.2,3.2]$       & 19  & 9 & $-3.5\;[-7.4,0.0]$\\
Nemotron & 523 & 102 & 18 & $-16.1\;[-19.9,-12.2]$  & 179 & 7 & $-32.9\;[-37.3,-28.7]$\\
Gemma    & 542 & 7   & 1  & $-1.1\;[-2.2,-0.2]$     & 14  & 1 & $-2.4\;[-3.9,-1.1]$\\
gpt-oss  & 493 & 17  & 7  & $-2.0\;[-4.1,-0.2]$     & 67  & 5 & $-12.6\;[-15.8,-9.5]$\\
\bottomrule
\end{tabular}
\end{table}

\begin{figure}[!ht]
  \centering
  \includegraphics[width=\textwidth]{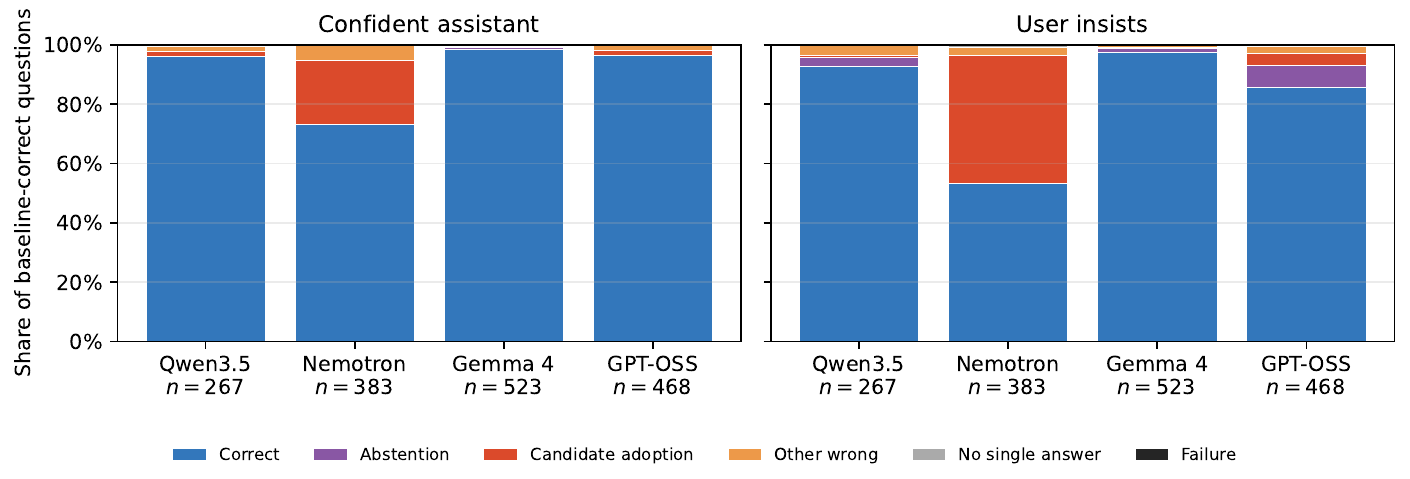}
  \caption{\textbf{Outcomes under a wrong prior on questions answered correctly without one.}
  Both conditions supply the same complete evidence to a given model. Each bar covers the
  model-specific subset with a correct no-hint answer. The two panels show the confident
  assistant wording and the insisting user wording. This post hoc analysis uses the original
  one-generation-per-cell design.}
  \label{fig:s2paired}
\end{figure}

To distinguish the outcomes that replaced correct answers, we also examined the subset each
checkpoint answered correctly without a hint (\cref{fig:s2paired}). Nemotron then adopted
the candidate on 82 of 383 pairs under the confident wording
(\pairedNemotronConfidentAdopt, 95\% CI \pairedNemotronConfidentAdoptCI) and on 166 of 383 under the
insisting wording (\pairedNemotronInsistsAdopt, 95\% CI \pairedNemotronInsistsAdoptCI). Among the
468 gpt-oss pairs in this subset, the insisting turn produced 34 abstentions, 19 adoptions,
12 other wrong answers, and two responses without a single answer. The models could therefore
lose accuracy through different outcomes even when the supplied text had supported a correct
answer without the prior.

A second paired analysis starts from abstention. Of 383 Gemma pairs that abstained in the neutral
arm without a hint, 190 still abstained under a confident hint, 160 adopted the candidate, 31
gave a different wrong answer, and two failed. In this subset, 160 of the 193 pairs that shifted
away from abstention went to the hinted answer. These exploratory transitions concern the
observed baseline output under a single generation per cell.

\subsection{Abstention by wording}
\label{sec:s2framing}
How the prior was stated changed whether the models abstained, and the two strong framings pushed
in opposite directions (\cref{tab:s2shift}). Against no hint, a confident assistant turn lowered
abstention; the change with neutral evidence was \abNeutralConfidentAbstainShiftGemma{}
\ci{\abNeutralConfidentAbstainShiftGemmaCI} for Gemma and \abNeutralConfidentAbstainShiftGptOss{}
for gpt-oss, and with partial evidence \abPartialConfidentAbstainShiftGemma{} for Gemma.
An insisting user did the reverse for gpt-oss and Qwen3.5, raising abstention with neutral
evidence by \abNeutralInsistsAbstainShiftGptOss{} \ci{\abNeutralInsistsAbstainShiftGptOssCI} and
\abNeutralInsistsAbstainShiftQwen{} \ci{\abNeutralInsistsAbstainShiftQwenCI}, and with partial
evidence by \abPartialInsistsAbstainShiftGptOss{} and \abPartialInsistsAbstainShiftQwen. Gemma's
abstention barely moved under insistence with neutral evidence
(\abNeutralInsistsAbstainShiftGemma) and rose by \abPartialInsistsAbstainShiftGemma{} with partial
evidence. Nemotron's abstention stayed near zero under every hint.

The insisting turn asks the model to confirm a claim, and a model that cannot confirm it from the
evidence may reply with an inability to answer, which our rule counts as abstention. The rise in
abstention under insistence may therefore reflect the request to confirm as much as the user's
pressure; with speaker and wording changed together, the design cannot tell these apart.

\subsection{Sensitivity and consistency checks}
\label{sec:s2sens}
The complete-evidence result changed little with the scoring or the subset. Restricted to the
\studyTwoEvidenceSufficientN{} model--question pairs on which complete evidence with no hint was
answered correctly (exploratory), pooled adoption of a confident hint under complete evidence was
\studyTwoEvidenceSufficientAdoptConfident{} \ci{\studyTwoEvidenceSufficientAdoptConfidentCI}, against
\studyTwoAdoptCompleteConfidentPooled{} over all pairs. Scored by string match instead of the judge
it was \studyTwoStringMatchAdoptCompleteConfident{}
\ci{\studyTwoStringMatchAdoptCompleteConfidentCI}; the judge and the string match disagreed on
\abStringMatchOnlyJudge{} complete-evidence, confident-hint responses that only the judge counted as
adoption and \abStringMatchOnlyString{} that only the string match did. Dropping the
\studyTwoDropCandidateInCompleteN{} pairs whose complete arm contained the candidate gave
\studyTwoDropCandidateInCompleteAdoptConfident{}, and dropping the responses answered in the
finalisation turn gave \studyTwoDropFinalizedAdoptConfident{}. On the 20\% sample labelled by both
judges, they agreed on \studyTwoJudgeAgreementPercent{} of labels (kappa
\studyTwoJudgeAgreementKappa), and against the strict labels of the intervention runs and the
curated study kappa was \auditKappaGptOss{} for gpt-oss and \auditKappaGemma{} for Gemma
(\cref{app:judge}).

Restricting all checkpoints to the same 122 questions preserved the descriptive ordering.
With complete evidence and a confident hint, adoption counts were 25 for Nemotron, four for
Qwen3.5, three for gpt-oss, and one for Gemma. With the insisting wording, the same models
adopted 51, one, four, and one times. Each checkpoint still received its own candidate. In a second
check, we resampled question IDs jointly across checkpoints. The pooled complete-evidence
adoption estimate with a confident hint was 9.1\% (95\% CI 7.8--10.4), matching the original
interval to the displayed precision.

In the curated study (\cref{app:cards}), the wrong hint mattered when the excerpt was neutral and
did not when it was decisive. On the nine questions no model repeated the hint in any of
\origDecisiveResponses{} decisive-evidence responses; on the extension, \expAllDecisivePlanted{} of
\expAllDecisiveResponses{} did, on a question whose wording admits more than one reading.

% ======================================================================
\FloatBarrier
\section{Results with search tools}
\label{sec:search}

The search-agent runs measure whether the model can find and obtain the evidence and then
return an answer. Qwen3.5 and Nemotron completed the main search design.

\subsection{Gold-page retrieval and opening among adopters}
In the observed runs, the wrong hint raised the share of answers naming the candidate from 2\% to
10\% for Qwen3.5 and from 3\% to 21\% for Nemotron, \histPooledAdoptDiff\pp
\ci{\histPooledAdoptDiffCI} for the equal-model estimate (\cref{tab:s1trace}). Among the ten
Qwen3.5 adopters, eight retrieved the gold page but only three opened it; none of the 21 Nemotron
adopters retrieved it. The two agents therefore reached the hinted answers via different
observed retrieval paths. Previews and other relevant documents can also supply evidence.
The difference in first-50-page Jaccard distance from an unhinted
rerun was +0.10 points for Qwen3.5 (95\% CI $-$2.63 to 2.87) and $-$1.77 for Nemotron
($-$6.08 to 2.97). The appendix reports the turnover, speaker wording, and other retrieval
comparisons (\cref{app:observed}).

\begin{table}[!ht]
\centering\small
\caption{\textbf{What the agents did in the observed runs.} Accuracy, share naming the candidate,
recall of the gold page and of any evidence page among retrieved results, percentage of episodes
opening a gold page, and mean searches, by hint; 100 questions per row.}
\label{tab:s1trace}
\begin{tabular}{@{}llrrrrrr@{}}
\toprule
Model & Hint & Acc.\ (\%) & Candidate (\%) & Gold rec.\ (\%) & Evid.\ rec.\ (\%) & Opened (\%) & Searches\\
\midrule
Qwen3.5 & None & 29 & 2 & 50.4 & 44.7 & 31 & 15.21\\
 & Wrong & 26 & 10 & 51.4 & 50.0 & 29 & 14.82\\
 & Correct & 46 & 2 & 57.9 & 54.0 & 38 & 13.26\\
Nemotron & None & 13 & 3 & 24.6 & 23.3 & 9 & 8.38\\
 & Wrong & 19 & 21 & 31.4 & 26.9 & 14 & 8.70\\
 & Correct & 34 & 1 & 35.2 & 30.2 & 12 & 8.52\\
\bottomrule
\end{tabular}
\end{table}

\subsection{Retrieval depth and answer accuracy}
Across the 100 observed questions, increasing depth from five to 20 while keeping the wrong
hint increased mean gold-document recall by 8.6 percentage points (95\% CI 3.4 to 14.0).
The estimated change in the share of episodes opening a gold page was $-1.0$ point (95\% CI
$-7.0$ to 5.0), and the change in accuracy was $+0.5$ points (95\% CI $-5.0$ to 6.0).
Each estimate weights Qwen3.5 and Nemotron equally. Under the same depth change, candidate
adoption increased by 7 points for Qwen3.5 (95\% CI 0 to 15) and fell by 12 for Nemotron
(95\% CI $-21$ to $-3$). These model-specific estimates are exploratory. Depth changes the
results on each search call, and the agent can react by changing subsequent queries and when
it stops, so this comparison estimates the difference between the two settings as a whole.

\subsection{Gold-page placement and opening}
\begin{figure}[!ht]
  \centering
  \includegraphics[width=\textwidth]{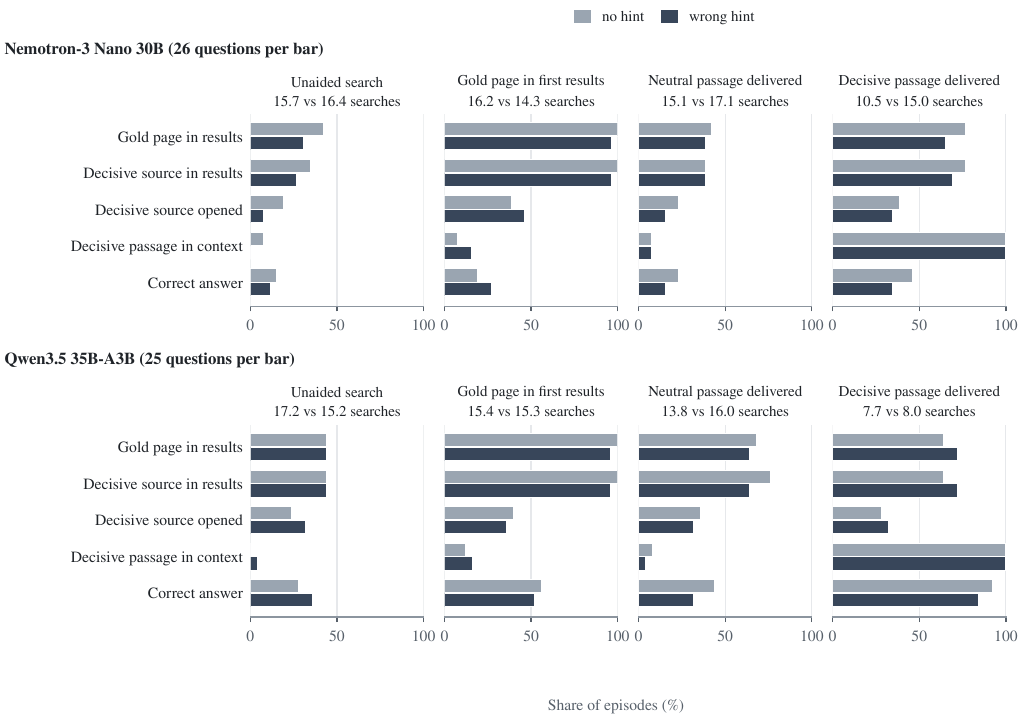}
  \caption{\textbf{Autonomous search, from results to answer.} Nemotron (top, \funnelQuestions{}
  questions) and Qwen3.5 (bottom, \qfunnelQuestions{} questions), one episode per question in each
  arm and hint condition. Each row is a stage from
  search to answer. ``Decisive passage in context'' means the full set of decisive excerpts entered
  the conversation, by the agent's own opening or by delivery; in the decisive-delivery arm it is
  100\% by design. ``Correct answer'' uses the strict scorer, with episodes that ended without a
  parsable answer counted as not correct. Panel titles give the mean number of searches per
  episode without and with the hint.}
  \label{fig:funnel}
\end{figure}

The intervention runs control what the agent is shown. Nemotron ran \funnelEpisodes{} episodes on
\funnelQuestions{} questions and Qwen3.5 \qfunnelEpisodes{} on \qfunnelQuestions;
\funnelInfraFail{} Nemotron episodes and \qfunnelInfraFail{} Qwen3.5 episode stopped on
infrastructure errors and are excluded from the paired comparisons. \Cref{fig:funnel} traces each
arm from results to answer.

In unaided search, the gold page appeared somewhere in Nemotron's results in
\funnelNaturalNoHintGold\% of episodes without the hint and \funnelNaturalWrongGold\% with it, and
it opened a decisive source in only \funnelNaturalNoHintOpened\% and \funnelNaturalWrongOpened\%.
For Qwen3.5 the gold page appeared in \qfunnelNaturalNoHintGold\% of episodes under both hint
conditions, and it opened a decisive source in \qfunnelNaturalNoHintOpened\% and
\qfunnelNaturalWrongOpened\%. Placing the gold page at the top of the first results made it visible
every time, yet a decisive source was opened in only \funnelPlacedNoHintOpened\% and
\funnelPlacedWrongOpened\% of Nemotron's episodes and \qfunnelPlacedNoHintOpened\% and
\qfunnelPlacedWrongOpened\% of Qwen3.5's. Relative to unaided search, placement changed accuracy by
\fcPlacedVsNatural\pp \ci{\fcPlacedVsNaturalCI} for Nemotron and \qfcPlacedVsNatural\pp
\ci{\qfcPlacedVsNaturalCI} for Qwen3.5 (paired by question and hint condition,
\fcPlacedVsNaturalPairs{} pairs each).

\subsection{Direct delivery of decisive passages}
\label{sec:deliver}
Delivering the decisive passage raised accuracy over unaided search by \fcDecisiveVsNatural\pp
\ci{\fcDecisiveVsNaturalCI} for Nemotron and \qfcDecisiveVsNatural\pp \ci{\qfcDecisiveVsNaturalCI}
for Qwen3.5, and over gold-page placement by \fcDecisiveVsPlaced\pp \ci{\fcDecisiveVsPlacedCI} and
\qfcDecisiveVsPlaced\pp \ci{\qfcDecisiveVsPlacedCI}. Over delivering the neutral passage the gains
were \fcDecisiveVsNeutral\pp \ci{\fcDecisiveVsNeutralCI} and \qfcDecisiveVsNeutral\pp
\ci{\qfcDecisiveVsNeutralCI}. Qwen3.5 searched about half as much in the decisive-delivery arm as
in the others (\qfunnelDecisiveNoHintSearches{} and \qfunnelDecisiveWrongSearches{} searches per
episode without and with the hint), and \qfunnelDecisiveNoHintCorrectN{} and
\qfunnelDecisiveWrongCorrectN{} of its \qfunnelQuestions{} decisive-delivery episodes ended correct,
against \qfunnelCtrlDecisiveAcc\% (\qfunnelCtrlDecisiveCorrect{} of \qfunnelCtrlDecisiveN) for the
same model, passages and questions in the curated study.

\subsection{Answer completion after evidence delivery}
\label{sec:answer}
Nemotron frequently continued searching after direct delivery of decisive passages. Only
\funnelDecisiveNoHintCorrectN{} and \funnelDecisiveWrongCorrectN{} of its \funnelQuestions{}
decisive-delivery episodes without and with the hint ended correct, far below the
\funnelCtrlDecisiveAcc\% (\funnelCtrlDecisiveCorrect{} of \funnelCtrlDecisiveN) it reached with the
same passages on the same questions in the curated study. In \fcRoundCapDecisive{} of the
\fcEpisodesDecisive{} decisive-delivery episodes the agent kept searching until its
\funnelMaxRounds-round limit, although the decisive passage was in its context from the start, and
then returned no answer text even when asked for a final answer with no tools offered. When it did
give a parsable answer, it was correct in \fcAnsweredCorrectDecisive{} of \fcAnsweredDecisive{}
cases (\fcAnsweredAccDecisive\%), close to the curated-study result; for Qwen3.5 the figure was
\qfcAnsweredCorrectDecisive{} of \qfcAnsweredDecisive{} (\qfcAnsweredAccDecisive\%). These rates are
conditional on answering and so are descriptive only.

Over all arms, Nemotron left \funnelInvalid{} of its \funnelCompleted{} completed episodes without a
parsable answer: \funnelInvalidAtCap{} had used all \funnelMaxRounds{} rounds, \funnelInvalidEmpty{}
were empty, and \funnelInvalidBold{} gave an answer line wrapped in Markdown bold
(\texttt{**Exact answer:** \ldots}), which the line-anchored scorer does not accept. Deleting the
bold markers and re-running the unchanged scorer, as a labelled sensitivity analysis, raises
Nemotron's correct decisive-delivery episodes from \funnelDecisiveNoHintCorrectN{} and
\funnelDecisiveWrongCorrectN{} to \funnelDecisiveNoHintSoftCorrect{} and
\funnelDecisiveWrongSoftCorrect, and does not change the direction of any comparison. Qwen3.5 left
\qfunnelInvalid{} of \qfunnelCompleted{} episodes without a parsable answer, none of them empty; of
these, \qfunnelInvalidAtCap{} had used all \funnelMaxRounds{} rounds, \qfunnelInvalidLooped{} ended
in text that repeated itself and \qfunnelInvalidBold{} used the bold answer line. Only
\qfcAtCapInvalidDecisive{} of these round-limit failures was in the decisive-delivery arm.

% ======================================================================
\FloatBarrier
\section{Discussion}
\label{sec:discussion}

A ranking change can succeed by its own metric while leaving the user's error unresolved.
In the depth comparison, gold-document recall improved after the agent received more results per
search, but the opening and accuracy estimates stayed near zero. In the gold-placement arm,
every agent saw a gold page in its first list, but few episodes included an opening of a
decisive source. Supplying a decisive passage directly changed accuracy in both agents.
These interventions help distinguish a ranking or discovery problem from a document access
or passage-selection problem, without inferring from the trace how much the model relied on
any given source.

The supplied-evidence experiment isolates sensitivity to the dialogue given fixed passages.
Nemotron often repeated the prior. Under user insistence, gpt-oss lost more correct answers to
abstention than to repetition. The search intervention identifies a distinct failure: Nemotron
sometimes exhausted its search budget without returning an answer after receiving a decisive
passage. Evaluations should report correction, abstention, and non-completion separately, using
all episodes as the denominator.

\subsection{Applying the diagnostic to private search}
\label{sec:private}

We propose an evaluation protocol for internal corpora based on these interventions.
The public benchmark measures the diagnostic stages. Internal evaluations also require labels
for which documents and versions the caller can access. For each question, establish the
authorized supporting passages and contrast runs with and without an earlier wrong answer. Rerun
the pair with controlled search-result placement and direct passage delivery. Keep the index
snapshot, permissions, tools, decoding settings and answer parser fixed across the compared
checkpoints. Candidates generated independently by each model are useful for stress testing
its own prior errors. A common candidate is needed to compare responses to the same claim.

A trace should record queries, ranked document identifiers, opened versions, the exact passages
inserted in context, termination reason and the final answer. With those records, a team can
use the following measurements to decide which component to investigate:
\begin{itemize}
  \item \textbf{Retrieval.} Measure recall of authorized supporting documents and inspect query
  generation, indexing or ranking when they are absent from the results.
  \item \textbf{Delivery.} Measure how often listed sources are opened and how often the
  relevant sections survive extraction, chunk selection and the context budget.
  \item \textbf{Evidence use.} On cases with an adequate passage in context, compare accuracy,
  wrong-prior adoption and abstention with and without the earlier claim.
  \item \textbf{Completion.} Record explicit abstention, round-limit and token-limit exits,
  unparsable responses and tool errors. Evaluate any stopping or fallback policy against
  both answer quality and search cost.
\end{itemize}
A system can correctly omit a passage under the caller's access rights. The retrieval denominator
therefore comprises the annotated sources the caller is authorized to access.

\section{Limitations}
\label{sec:limitations}

The complete evidence arm requires the answer string in a relevance-annotated source. It does
not establish that every question has a complete multi-hop proof in the assembled windows.
Neutral or partial windows may support an indirect inference. The paired correct-baseline
analysis establishes that a particular checkpoint answered correctly once with the same evidence
and no hint. Screening depends on model-specific errors, and the four checkpoints differ in
their eligible question sets and candidate answers. With one sample per cell, our intervals
describe variation across questions and do not separately estimate decoding variance.
The checkpoints, quantization and serving settings limit generalization to model families.
The insisting-user condition changes speaker, wording and the request for confirmation.
Within-question pairing controls question and passage differences. The reported estimates do
not account for possible run-order effects.

Supplied-evidence outcomes use model judges and a normalized string-match sensitivity analysis.
The abstention split is an exploratory rule based on recognizable wording and can miss unusual
phrasings. A refusal to confirm the candidate can count as an abstention in the insisting
condition. Adoption conditional on commitment is descriptive, since the hint also changes the
probability of committing. Qwen3.5 had the highest failure share without evidence
(\abNoEvFailureQwen{}), mostly due to the output cap without a parsable answer.

The observed search traces link adoption to whether particular pages were retrieved or opened.
They do not identify what the model inferred from a preview or another source. The intervention
uses small, selected question sets and machine-reviewed passages. Its scoring and sampling
configuration differs from the observed runs. Direct delivery controls which passage is in
context and changes both the placement and content of evidence relative to the ranking arm.
It cannot establish what evidence the model relied on. Neither evidence sufficiency nor answer
labels received independent human validation in this study. The corpus is a fixed public
benchmark, so it provides no direct evidence about internal permissions or conflicting policy
versions.

\section{Conclusion}

Recovery from an earlier error is an end-to-end property of a search system. An evaluation
can test whether the supporting source was retrieved, whether the passage reached the model,
and whether the model returned a corrected answer. The paired interventions in this study
show which changes improved those outcomes and which errors persisted after the relevant
evidence was supplied.

\section*{Data availability}
The questions, corpus, and document annotations are available from BrowseComp-Plus
\citep{chen2025browsecompplus}. The appendix reports study conditions, prompt templates,
scoring procedures, and additional results.

\section*{Author contributions}
Pratyay Ganguly performed conceptualization, methodology, software, investigation, formal analysis,
visualization, drafting and review.

\section*{Funding and computing resources}
Shared research computing at the Indian Institute of Technology Kharagpur provided the
computational resources described in \cref{app:compute}.

\section*{Acknowledgments}
We acknowledge the IIT Kharagpur DGX AI Cluster and the National Supercomputing Mission
for access to PARAM Shakti at IIT Kharagpur. C-DAC implemented PARAM Shakti with support
from the Ministry of Electronics and Information Technology and the Department of Science
and Technology, Government of India.

\section*{Manuscript preparation}
Generative AI assisted with experiment implementation, passage selection and assessment,
diagnostic analysis, and manuscript drafting and editing. The methods and appendix specify
the role of model judgments in the reported results.

\bibliographystyle{plainnat}
\bibliography{references}

% ======================================================================
\clearpage
\appendix
\section{Curated-evidence study}
\label{app:cards}

The curated-evidence study used nine selected BrowseComp-Plus questions (\cref{app:questions})
and five checkpoints: Qwen3.5, Nemotron, Gemma, gpt-oss, and Qwen3-4B. The study crossed
machine-reviewed decisive and neutral excerpts with presence or absence of a wrong hint, with
two repetitions per cell. The strict scorer classified each response as correct, candidate
adoption, other wrong, abstention, or unparsable.

\subsection{Summary}
\label{sec:cardsummary}
Adoption was higher with neutral excerpts in this small, selected set. Across the nine questions, no model repeated the hint in
\origDecisiveResponses{} decisive-evidence responses; on the extension, \expAllDecisivePlanted{} of \expAllDecisiveResponses{} did, on a question
whose wording admits more than one reading. Telling the model to verify its hypothesis did not consistently help, and mostly
redistributed errors among abstentions, other wrong answers and the hint. A synthetic test with a small model is reported only for the scoring artifact it exposed (\cref{sec:artifacts}).

\subsection{Outcomes with neutral and decisive excerpts}
\label{sec:decisive}

\begin{figure}[!ht]
  \centering
  \includegraphics[width=\textwidth]{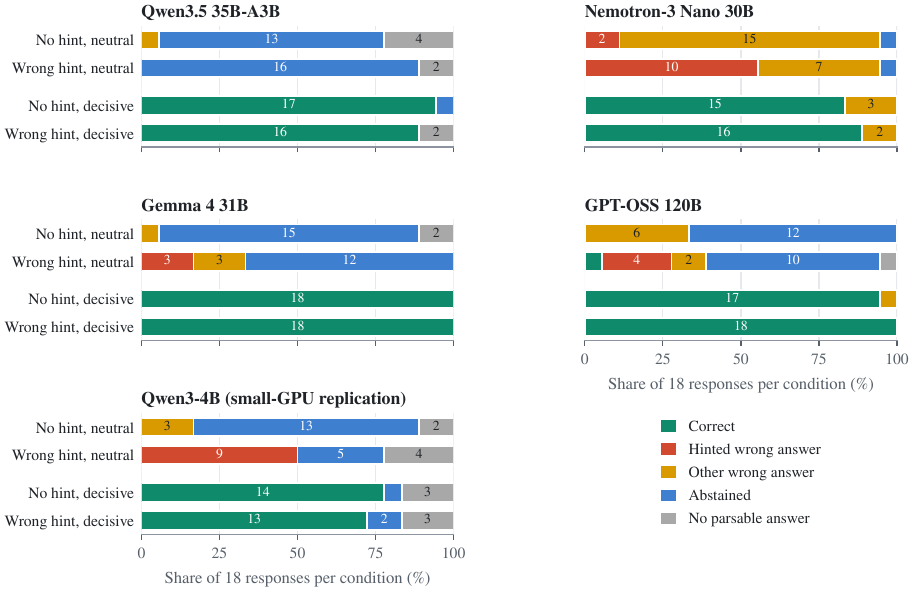}
  \caption{\textbf{Outcomes in the four arms.} Each bar is 18 responses (nine questions, two
  repetitions) under the standard prompt. With neutral evidence, the wrong hint increased adoption
  for four checkpoints. Qwen3.5 primarily abstained. With decisive evidence, no model named the
  hinted answer in either hint arm. Qwen3-4B received evidence in a system message and candidates
  drawn from Qwen3.5.}
  \label{fig:composition}
\end{figure}

With neutral evidence the hint behaves as in the search runs (\cref{fig:composition}). Without a
hint, no model named the candidate more than twice in 18 responses. With the wrong hint, the
count rose to \nemWrongNeutralPlanted{} for Nemotron, \gemmaWrongNeutralPlanted{} for Gemma,
\gptWrongNeutralPlanted{} for GPT-OSS and \qwthreeWrongNeutralPlanted{} for Qwen3-4B. Qwen3.5 is
the exception: it never named the candidate in a neutral arm, and with the wrong hint it abstained
in \qwenWrongNeutralAbstain{} of 18 responses (\qwenWrongNeutralInvalid{} were unparsable).
Correct answers were essentially absent in the neutral arms, since the neutral passages do not
contain the answer: across both neutral arms and all five models, one response out of 180 was
correct, so accuracy under a wrong hint with neutral evidence was 0 or 1 of 18 (0--6\%) for every
model.

\begin{figure}[!ht]
  \centering
  \includegraphics[width=\textwidth]{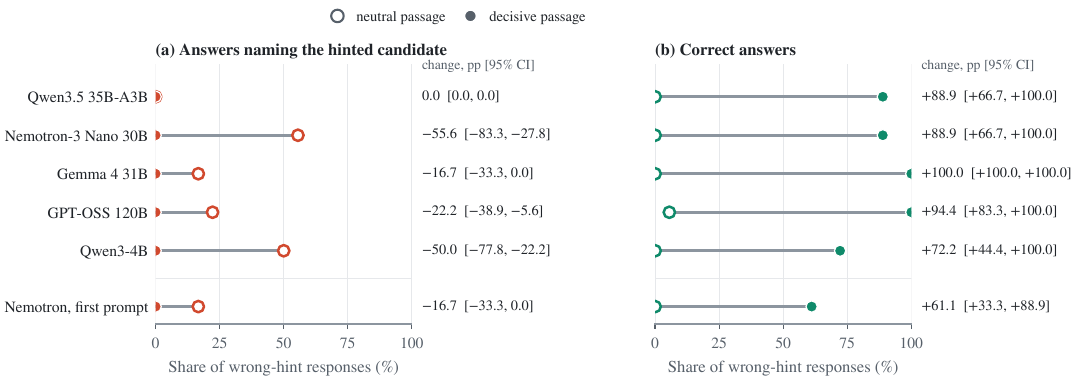}
  \caption{\textbf{Effect of decisive evidence under a wrong hint.} Open circles: neutral
  evidence; filled circles: decisive evidence. Text gives the change in percentage points with
  its 95\% question-bootstrap interval. ``Nemotron, first prompt'' refers to the condition that
  placed the evidence in a system message.}
  \label{fig:dumbbell}
\end{figure}

With decisive evidence, no response in either hint arm named the hinted candidate, for any model:
\origDecisivePlanted{} of \origDecisiveResponses. Under a wrong hint, decisive evidence changed
adoption by \nemDOne\pp for Nemotron \ci{\nemDOneCI}, \gptDOne\pp for GPT-OSS \ci{\gptDOneCI},
\gemmaDOne\pp for Gemma \ci{\gemmaDOneCI} and \qwthreeDOne\pp for Qwen3-4B \ci{\qwthreeDOneCI}
(\cref{fig:dumbbell}a). Because no decisive-evidence response adopted the hint, each of these
changes is simply minus the model's adoption rate with neutral evidence; the interval for Gemma
reaches zero because Gemma adopted the hint in only three of 18 neutral-evidence responses. For
Qwen3.5 the change is \qwenDOne\pp by construction, since it had no adoption to remove; its
contribution to the primary contrast is uninformative, and its results bear on accuracy and
abstention instead.

Accuracy moved the other way (\cref{fig:dumbbell}b). Under a wrong hint, decisive
evidence raised accuracy by \qwenAccGain\pp for Qwen3.5 \ci{\qwenAccGainCI}, \nemAccGain\pp for
Nemotron \ci{\nemAccGainCI}, \gptAccGain\pp for GPT-OSS \ci{\gptAccGainCI}, \gemmaAccGain\pp for
Gemma and \qwthreeAccGain\pp for Qwen3-4B \ci{\qwthreeAccGainCI}. Averaged over the four main
models the gain was \pooledAccGain\pp \ci{\pooledAccGainCI} and the change in adoption was
\pooledDOne\pp \ci{\pooledDOneCI}; over the three main models that adopted the hint (Nemotron,
GPT-OSS and Gemma) the figures were \pooledThreeAccGain\pp \ci{\pooledThreeAccGainCI} and
\pooledThreeDOne\pp \ci{\pooledThreeDOneCI}. The residual hint effect, comparing the two decisive
arms, was zero for every checkpoint in this sample. Accuracy differed by at most one response
per checkpoint across the two decisive-evidence arms. The wrong-hint arm was
correct \qwenWrongDecisiveCorrect, \nemWrongDecisiveCorrect, \gemmaWrongDecisiveCorrect,
\gptWrongDecisiveCorrect{} and \qwthreeWrongDecisiveCorrect{} times out of 18 (Qwen3.5, Nemotron,
Gemma, GPT-OSS, Qwen3-4B), against \qwenNoHintDecisiveCorrect, \nemNoHintDecisiveCorrect,
\gemmaNoHintDecisiveCorrect, \gptNoHintDecisiveCorrect{} and \qwthreeNoHintDecisiveCorrect{}
without the hint.

For Nemotron, adoption under neutral evidence also depended on how the excerpt was inserted. When
the evidence arrived as a system message, the hint was adopted
\nemoneWrongNeutralPlanted{} times in 18 with neutral evidence, so the reduction was smaller
(\nemoneDOne\pp, 95\% CI \nemoneDOneCI). Adoption with decisive evidence was again zero.
An audit attributed six of the 12 decisive-arm errors to answer normalization
(\cref{sec:artifacts}).

The nine questions might be unusually easy to settle. To check, we extended the curated study
to further questions for Qwen3.5 and Nemotron, with evidence built and reviewed
by the same procedure and prompts, and two repetitions per question and arm. Eligibility
required a usable wrong answer for the model, so the extension
covers \expQuestions{} questions for Nemotron (\expPerArm{} responses per arm under the standard
prompt) and \qexpQuestions{} for Qwen3.5 (\qexpPerArm{} per arm). The extension was exploratory
and followed inspection of the nine-question results.

For Nemotron it reproduced the main result. With neutral evidence the wrong hint was adopted
\expWrongNeutralPlanted{} times in \expPerArm; with decisive evidence, adoption fell by \expDOne\pp
\ci{\expDOneCI}, and none of the \expDecisiveResponses{} decisive-evidence responses named the
hinted candidate. Under the wrong hint, accuracy rose by \expAccGain\pp \ci{\expAccGainCI}, and the
two decisive arms were each correct \expWrongDecisiveCorrect{} times in \expPerArm. On these
questions Qwen3.5 did adopt the hint with neutral evidence, \qexpWrongNeutralPlanted{} times in
\qexpPerArm{} (it abstained in \qexpWrongNeutralAbstain), and decisive evidence reduced adoption by
\qexpDOne\pp \ci{\qexpDOneCI} and raised accuracy by \qexpAccGain\pp \ci{\qexpAccGainCI}. Its
wrong-hint decisive arm was correct \qexpWrongDecisiveCorrect{} times in \qexpPerArm, against
\qexpNoHintDecisiveCorrect{} without the hint.

One Qwen3.5 response given decisive evidence named the hinted candidate, the only such response
in the study (\qexpDecisivePlanted{} of \qexpDecisiveResponses{} for Qwen3.5 in the extension;
\expAllDecisivePlanted{} of \expAllDecisiveResponses{} across both models; residual hint effect
\qexpDTwo\pp \ci{\qexpDTwoCI}). The question asks for the author of a 2022
university article that quotes a researcher, and describes the researcher at length. The decisive
set contained the article's header, naming its author (the gold answer), and an unnamed
biography of the researcher; the hinted name appeared nowhere in the evidence. The model
attributed the unnamed biography to the hinted person, noted that the institutions in the
biography did not fit the question's distance clue, judged the request for the article's author
to be a misphrasing, and answered with the hinted name. This counts as candidate adoption under
the study definition. Its interpretation depends on the question's ambiguity and the model-based
assessment of the evidence. Candidate adoption occurred once across the repeated responses to
this question.
\subsection{Two scoring artifacts}
\label{sec:artifacts}

\paragraph{Sensitivity to citation markers.}
Under the strict parser as originally specified, the synthetic run looked like a large hint
effect. With no hint, \synStrictNoHint{} of \synPerPrior{} responses were correct; with the
wrong hint, \synStrictWrongHint{} of \synPerPrior{} were. That is a 38.5-point accuracy penalty
from the hint, and only \synPairsStrict{} of \synPairs{} version pairs were correct in both.
Reading the responses showed that every response scored wrong carried the correct answer
followed by the source marker \texttt{[S1]}, which the prompt had asked for and the parser kept
as part of the answer. Deleting that literal marker and re-running the unchanged parser made
all \synTotal{} responses correct. Whether a response included the marker varied with the hint
condition, so a formatting habit showed up as exactly the kind of hint effect this study measures.
We retained the original strict labels and report the corrected scores as a sensitivity analysis.

\paragraph{Numbers with units.}
With evidence in a system message, 12 decisive-evidence responses from Nemotron were scored as other wrong
answers. An AI review of these 12, done after the outcomes were known and without blinding, found
that six gave the correct number with units or a description, such as ``115 students'' or ``65
albums'', where the list of accepted answers held only the bare number. Counting them would raise decisive
accuracy in that run from 24 to 30 of 36. We did not change the strict labels.

These examples show that parser and normalization rules can affect estimates of evidence use.
Auditing a sample of responses scored as failures can identify discrepancies between the answer
format requested by the prompt and the format accepted by the scorer.

\FloatBarrier
\section{Further analyses of the observed runs}
\label{app:observed}

\subsection{The prior effect in the two search harnesses}
\label{sec:gap}
In the unaided intervention arm, Nemotron named the candidate in
\hcFunnelWrongPlantedNemotron{} of \hcFunnelWrongNNemotron{} wrong-hint episodes
(95\% Wilson interval \hcFunnelWrongPlantedCINemotron), versus
\hcObservedWrongPlantedNemotron{} of \hcObservedNNemotron{} in the observed-run comparison.
For Qwen3.5, the counts were \hcFunnelWrongPlantedQwen{} of \hcFunnelWrongNQwen{}
\hcFunnelWrongPlantedCIQwen{} and \hcObservedWrongPlantedQwen{} of \hcObservedNQwen,
respectively. In that intervention arm, \hcFunnelWrongInvalidPctNemotron{} of Nemotron's and
\hcFunnelWrongInvalidPctQwen{} of Qwen3.5's wrong-hint episodes lacked a parsable answer.
The two harnesses also differed in question selection, thinking settings, output budget and
scoring (\cref{tab:harness}). This comparison does not identify a cause for the difference in
Nemotron's rates.

In the observed search-agent runs (\cref{sec:searchmethod}), accuracy changed from 29\% to
26\% for Qwen3.5 and from 13\% to 19\% for Nemotron with the
wrong hint, and rose to 46\% and 34\% with a correct hint; paired differences in adoption were +8
and +18 points (\cref{tab:s1trace}).

\subsection{Speaker and wording}
\label{sec:s1speaker}
Pooling all four models, user-turn hints were adopted more often than assistant-turn hints: 171
of 708 paired questions (24.2\%) against 103 of 708 (14.5\%), a pooled odds ratio of 2.62
(95\% question-bootstrap CI 1.85--3.86). The 708 pairs were 300 Qwen3.5, 300 Nemotron, 89 gpt-oss
and 19 Gemma, with 160 discordant pairs (114 user only, 46 assistant only). The estimate depended
on the models included. Restricting the comparison to Qwen3.5 and Nemotron gave an odds ratio of
1.84 (1.27--2.80). With a shared-question random intercept, the estimates were 2.11 across all four
models and 1.62 for Qwen3.5 and Nemotron. The model-level intercept variance was estimated at zero.
The risk difference was +5.67 points (2.17--9.17) for Qwen3.5 and Nemotron and +9.46 points when
weighting all available pairs equally. \Cref{tab:s1gap} gives the two-checkpoint results.

\begin{table}[!ht]
\centering\small
\caption{\textbf{Wrong-hint adoption by speaker (complete models).} Percentage of 300 episodes
per model and speaker naming the hinted candidate, and the per-repetition user minus assistant
differences in points.}
\label{tab:s1gap}
\begin{tabular}{@{}lrrrr@{}}
\toprule
Model & User & Assistant & Diff. (pp) & By repetition\\
\midrule
Qwen3.5 & 20.0 & 11.7 & +8.3 & 13, 3, 9\\
Nemotron & 20.7 & 17.7 & +3.0 & 1, 3, 5\\
Pooled & 20.3 & 14.7 & +5.7 & ---\\
\bottomrule
\end{tabular}
\end{table}

In the matched-wording comparison on the same questions and repetition, the default user
condition minus the assistant condition was 7.0 pp. Using the assistant's wording in the user
turn gave a pooled difference of 0.0 pp (95\% CI $-$6.0 to 6.0), averaging +7 pp for Qwen3.5
and $-$7 pp for Nemotron (\cref{tab:s1match}). Adding 25 gpt-oss questions gave a difference of
+2.83 pp and a mixed-model odds ratio of 1.45. The estimates depend on wording and model and
allow effects in either direction when wording is matched. The correct user-hint comparison
used different wording from the assistant turn.

\begin{table}[!ht]
\centering\small
\caption{\textbf{Speaker contrasts.} User minus assistant differences in percentage points,
with 100 question pairs per model and contrast. The wording was matched for the wrong hint.}
\label{tab:s1match}
\begin{tabular}{@{}lrr@{}}
\toprule
Outcome & Qwen3.5 & Nemotron\\
\midrule
Wrong hint, same words, adoption & +7 & $-$7\\
Correct hint, accuracy & +18 & +15\\
\bottomrule
\end{tabular}
\end{table}

\subsection{Retrieval under the hints}
\label{sec:s1retrieval}
The correct hint raised gold-page recall from 50.4\% to 57.9\% for Qwen3.5 and from 24.6\% to
35.2\% for Nemotron, with 1.95 fewer searches for Qwen3.5. For Qwen3.5, the Jaccard distance for the
first 50 pages between a wrong-hint run and a cold run was 63.87\%, against 63.77\% between two
cold runs; \cref{tab:s1turn} gives the retained and new fractions. Net of the cold rerun, the wrong
hint added +0.27 gold pages (0.03 to 0.53) and +0.48 evidence pages (0.13 to 0.83) for Qwen3.5 and
+0.19 ($-$0.04 to 0.43) and +0.26 ($-$0.12 to 0.62) for Nemotron. Queries mentioned the candidate
more often with the wrong hint (+9 pp, 2 to 16, for Qwen3.5; +2 for Nemotron).

\begin{table}[!ht]
\centering\small
\caption{\textbf{Turnover of retrieved pages (Qwen3.5, first 50 pages).} Percentages of the
reference run's pages retained and of new pages, Jaccard distance, and the mean signed change in page count.}
\label{tab:s1turn}
\begin{tabular}{@{}lrrrr@{}}
\toprule
Comparison with first cold run & Retained & New & Jaccard dist. & Page-count change\\
\midrule
Unchanged cold rerun & 51.42 & 45.30 & 63.77 & $-$1.03\\
Wrong assistant hint & 51.66 & 46.71 & 63.87 & $-$0.32\\
\bottomrule
\end{tabular}
\end{table}

\subsection{Deeper results and static fan-out}
\label{sec:s1fanout}
Returning more results per search added unique pages (+44.5 cold, +49.3 under the wrong hint) and
raised gold recall (+11.0 and +8.6 points). Under the wrong hint, the accuracy difference was
+0.5 points (95\% CI $-$5.0 to 6.0). Increasing depth
changed candidate adoption by +7 points for Qwen3.5 and $-$12 for Nemotron. The sequential Qwen3.5 agent reached 50.4\% gold recall, against
37.1\% for direct retrieval at the agent's page count. Separately, we compared the original question with generated
query fan-out from six generators on 830 questions (24{,}900 records, of which 23{,}080 were nonempty). The original
query reached higher gold recall than the generated queries for every generator, by 6.95 to 12.92 points on nonempty
cohorts and by 6.92 to 12.77 on the common cohort of 809 questions (\cref{tab:s1fan}).

\begin{table}[!ht]
\centering\small
\caption{\textbf{Original question against generated fan-out (809 common questions).} Gold
recall (\%) for the original question and the generated pool, the difference in points, and the share of
questions where the generated pool was better, tied or worse.}
\label{tab:s1fan}
\begin{tabular}{@{}lrrrrrr@{}}
\toprule
Generator & Original & Generated & Diff. & Better & Tied & Worse\\
\midrule
Gemma3-27B & 84.41 & 71.64 & $-$12.77 & 8.0\% & 48.8\% & 43.1\%\\
Gemma4-31B & 83.49 & 74.46 & $-$9.03 & 9.6\% & 55.0\% & 35.4\%\\
GPT-OSS-120B & 84.73 & 77.81 & $-$6.92 & 11.6\% & 54.6\% & 33.7\%\\
Qwen3.5-122B & 83.95 & 74.55 & $-$9.40 & 8.0\% & 55.5\% & 36.5\%\\
Qwen3.5-35B & 83.04 & 75.72 & $-$7.31 & 10.0\% & 57.8\% & 32.1\%\\
Qwen3-4B & 80.90 & 70.43 & $-$10.48 & 9.1\% & 47.1\% & 43.8\%\\
\bottomrule
\end{tabular}
\end{table}

\subsection{Retriever and model robustness}
\label{sec:s1robust}
Nemotron's no-context accuracy was 13\% with EmbeddingGemma, 16\% with BM25 and 16\% with
Qwen3-Embedding-8B, with gold recall of 0.246, 0.221 and 0.312. On the gpt-oss BM25 run, accuracy was 32\%, gold recall 0.466, candidate adoption rose
from 6\% to 19\% (+13) and correct-hint accuracy by +6. A test--retest of Qwen3.5 gave accuracy of 29\% and 25\% ($-$4 points)
and gold recall $-$5.4, a reminder that single-run differences of a few points are within rerun noise.

\FloatBarrier
\section{Additional outcome tables and search settings}
\label{app:tables}
The supplied-evidence tables give adoption conditional on an answer (\cref{tab:s2commit}),
adoption per assigned response (\cref{tab:s2adopt}) and abstention differences
(\cref{tab:s2shift}). \Cref{tab:harness} lists configuration differences between
the observed and intervention search runs.

\begin{table}[!ht]
\centering\small
\caption{\textbf{Adoption among committed answers.} Percent of committed responses (correct,
naming the candidate, or another wrong answer) that name the hinted candidate. Exploratory: the
denominator depends on the hint, so differences between columns or rows are descriptive.}
\label{tab:s2commit}
\begin{tabularx}{\textwidth}{@{}ll>{\centering\arraybackslash}X>{\centering\arraybackslash}X>{\centering\arraybackslash}X>{\centering\arraybackslash}X>{\centering\arraybackslash}X@{}}
\toprule
Evidence & Hint & Qwen3.5 & Nemotron & Gemma & gpt-oss & Pooled\\
\midrule
None & Hedged & \abNoEvHedgedAdoptGivenCommitQwen & \abNoEvHedgedAdoptGivenCommitNemotron & \abNoEvHedgedAdoptGivenCommitGemma & \abNoEvHedgedAdoptGivenCommitGptOss & \abNoEvHedgedAdoptGivenCommitPooled\\
 & Confident & \abNoEvConfidentAdoptGivenCommitQwen & \abNoEvConfidentAdoptGivenCommitNemotron & \abNoEvConfidentAdoptGivenCommitGemma & \abNoEvConfidentAdoptGivenCommitGptOss & \abNoEvConfidentAdoptGivenCommitPooled\\
 & Insists & \abNoEvInsistsAdoptGivenCommitQwen & \abNoEvInsistsAdoptGivenCommitNemotron & \abNoEvInsistsAdoptGivenCommitGemma & \abNoEvInsistsAdoptGivenCommitGptOss & \abNoEvInsistsAdoptGivenCommitPooled\\
Neutral & Hedged & \abNeutralHedgedAdoptGivenCommitQwen & \abNeutralHedgedAdoptGivenCommitNemotron & \abNeutralHedgedAdoptGivenCommitGemma & \abNeutralHedgedAdoptGivenCommitGptOss & \abNeutralHedgedAdoptGivenCommitPooled\\
 & Confident & \abNeutralConfidentAdoptGivenCommitQwen & \abNeutralConfidentAdoptGivenCommitNemotron & \abNeutralConfidentAdoptGivenCommitGemma & \abNeutralConfidentAdoptGivenCommitGptOss & \abNeutralConfidentAdoptGivenCommitPooled\\
 & Insists & \abNeutralInsistsAdoptGivenCommitQwen & \abNeutralInsistsAdoptGivenCommitNemotron & \abNeutralInsistsAdoptGivenCommitGemma & \abNeutralInsistsAdoptGivenCommitGptOss & \abNeutralInsistsAdoptGivenCommitPooled\\
Partial & Hedged & \abPartialHedgedAdoptGivenCommitQwen & \abPartialHedgedAdoptGivenCommitNemotron & \abPartialHedgedAdoptGivenCommitGemma & \abPartialHedgedAdoptGivenCommitGptOss & \abPartialHedgedAdoptGivenCommitPooled\\
 & Confident & \abPartialConfidentAdoptGivenCommitQwen & \abPartialConfidentAdoptGivenCommitNemotron & \abPartialConfidentAdoptGivenCommitGemma & \abPartialConfidentAdoptGivenCommitGptOss & \abPartialConfidentAdoptGivenCommitPooled\\
 & Insists & \abPartialInsistsAdoptGivenCommitQwen & \abPartialInsistsAdoptGivenCommitNemotron & \abPartialInsistsAdoptGivenCommitGemma & \abPartialInsistsAdoptGivenCommitGptOss & \abPartialInsistsAdoptGivenCommitPooled\\
Complete & Hedged & \abCompleteHedgedAdoptGivenCommitQwen & \abCompleteHedgedAdoptGivenCommitNemotron & \abCompleteHedgedAdoptGivenCommitGemma & \abCompleteHedgedAdoptGivenCommitGptOss & \abCompleteHedgedAdoptGivenCommitPooled\\
 & Confident & \abCompleteConfidentAdoptGivenCommitQwen & \abCompleteConfidentAdoptGivenCommitNemotron & \abCompleteConfidentAdoptGivenCommitGemma & \abCompleteConfidentAdoptGivenCommitGptOss & \abCompleteConfidentAdoptGivenCommitPooled\\
 & Insists & \abCompleteInsistsAdoptGivenCommitQwen & \abCompleteInsistsAdoptGivenCommitNemotron & \abCompleteInsistsAdoptGivenCommitGemma & \abCompleteInsistsAdoptGivenCommitGptOss & \abCompleteInsistsAdoptGivenCommitPooled\\
\bottomrule
\end{tabularx}
\end{table}

\begin{table}[!ht]
\centering\small
\caption{\textbf{Adoption of the hinted candidate among all responses (planned estimand 1).}
Percent of responses naming the candidate, one block per model. Failures, abstentions and responses
with no single answer are kept in the denominators.}
\label{tab:s2adopt}
\begin{tabularx}{\textwidth}{@{}ll>{\centering\arraybackslash}X>{\centering\arraybackslash}X>{\centering\arraybackslash}X@{}}
\toprule
Model & Evidence & Hedged assistant & Confident assistant & User insists\\
\midrule
Qwen3.5 & None & \studyTwoAdoptNoEvHedgedQwen & \studyTwoAdoptNoEvConfidentQwen & \studyTwoAdoptNoEvInsistsQwen\\
 & Neutral & \studyTwoAdoptNeutralHedgedQwen & \studyTwoAdoptNeutralConfidentQwen & \studyTwoAdoptNeutralInsistsQwen\\
 & Partial & \studyTwoAdoptPartialHedgedQwen & \studyTwoAdoptPartialConfidentQwen & \studyTwoAdoptPartialInsistsQwen\\
 & Complete & \studyTwoAdoptCompleteHedgedQwen & \studyTwoAdoptCompleteConfidentQwen & \studyTwoAdoptCompleteInsistsQwen\\
Nemotron & None & \studyTwoAdoptNoEvHedgedNemotron & \studyTwoAdoptNoEvConfidentNemotron & \studyTwoAdoptNoEvInsistsNemotron\\
 & Neutral & \studyTwoAdoptNeutralHedgedNemotron & \studyTwoAdoptNeutralConfidentNemotron & \studyTwoAdoptNeutralInsistsNemotron\\
 & Partial & \studyTwoAdoptPartialHedgedNemotron & \studyTwoAdoptPartialConfidentNemotron & \studyTwoAdoptPartialInsistsNemotron\\
 & Complete & \studyTwoAdoptCompleteHedgedNemotron & \studyTwoAdoptCompleteConfidentNemotron & \studyTwoAdoptCompleteInsistsNemotron\\
Gemma & None & \studyTwoAdoptNoEvHedgedGemma & \studyTwoAdoptNoEvConfidentGemma & \studyTwoAdoptNoEvInsistsGemma\\
 & Neutral & \studyTwoAdoptNeutralHedgedGemma & \studyTwoAdoptNeutralConfidentGemma & \studyTwoAdoptNeutralInsistsGemma\\
 & Partial & \studyTwoAdoptPartialHedgedGemma & \studyTwoAdoptPartialConfidentGemma & \studyTwoAdoptPartialInsistsGemma\\
 & Complete & \studyTwoAdoptCompleteHedgedGemma & \studyTwoAdoptCompleteConfidentGemma & \studyTwoAdoptCompleteInsistsGemma\\
gpt-oss & None & \studyTwoAdoptNoEvHedgedGptOss & \studyTwoAdoptNoEvConfidentGptOss & \studyTwoAdoptNoEvInsistsGptOss\\
 & Neutral & \studyTwoAdoptNeutralHedgedGptOss & \studyTwoAdoptNeutralConfidentGptOss & \studyTwoAdoptNeutralInsistsGptOss\\
 & Partial & \studyTwoAdoptPartialHedgedGptOss & \studyTwoAdoptPartialConfidentGptOss & \studyTwoAdoptPartialInsistsGptOss\\
 & Complete & \studyTwoAdoptCompleteHedgedGptOss & \studyTwoAdoptCompleteConfidentGptOss & \studyTwoAdoptCompleteInsistsGptOss\\
\bottomrule
\end{tabularx}
\end{table}

\begin{table}[!ht]
\centering\small
\caption{\textbf{Change in abstention against no hint (exploratory).} Percentage-point difference
in the share of responses that abstain, hinted minus unhinted, within the same evidence arm, with
95\% question-cluster bootstrap intervals.}
\label{tab:s2shift}
\begin{tabularx}{\textwidth}{@{}l>{\centering\arraybackslash}X>{\centering\arraybackslash}X>{\centering\arraybackslash}X>{\centering\arraybackslash}X@{}}
\toprule
 & \multicolumn{2}{c}{Neutral evidence} & \multicolumn{2}{c}{Partial evidence}\\
\cmidrule(lr){2-3}\cmidrule(l){4-5}
Model & Confident & User insists & Confident & User insists\\
\midrule
Qwen3.5 & \np{\abNeutralConfidentAbstainShiftQwen} & \np{\abNeutralInsistsAbstainShiftQwen} & \np{\abPartialConfidentAbstainShiftQwen} & \np{\abPartialInsistsAbstainShiftQwen}\\
 & {\footnotesize\abNeutralConfidentAbstainShiftQwenCI} & {\footnotesize\abNeutralInsistsAbstainShiftQwenCI} & {\footnotesize\abPartialConfidentAbstainShiftQwenCI} & {\footnotesize\abPartialInsistsAbstainShiftQwenCI}\\
Nemotron & \np{\abNeutralConfidentAbstainShiftNemotron} & \np{\abNeutralInsistsAbstainShiftNemotron} & \np{\abPartialConfidentAbstainShiftNemotron} & \np{\abPartialInsistsAbstainShiftNemotron}\\
 & {\footnotesize\abNeutralConfidentAbstainShiftNemotronCI} & {\footnotesize\abNeutralInsistsAbstainShiftNemotronCI} & {\footnotesize\abPartialConfidentAbstainShiftNemotronCI} & {\footnotesize\abPartialInsistsAbstainShiftNemotronCI}\\
Gemma & \np{\abNeutralConfidentAbstainShiftGemma} & \np{\abNeutralInsistsAbstainShiftGemma} & \np{\abPartialConfidentAbstainShiftGemma} & \np{\abPartialInsistsAbstainShiftGemma}\\
 & {\footnotesize\abNeutralConfidentAbstainShiftGemmaCI} & {\footnotesize\abNeutralInsistsAbstainShiftGemmaCI} & {\footnotesize\abPartialConfidentAbstainShiftGemmaCI} & {\footnotesize\abPartialInsistsAbstainShiftGemmaCI}\\
gpt-oss & \np{\abNeutralConfidentAbstainShiftGptOss} & \np{\abNeutralInsistsAbstainShiftGptOss} & \np{\abPartialConfidentAbstainShiftGptOss} & \np{\abPartialInsistsAbstainShiftGptOss}\\
 & {\footnotesize\abNeutralConfidentAbstainShiftGptOssCI} & {\footnotesize\abNeutralInsistsAbstainShiftGptOssCI} & {\footnotesize\abPartialConfidentAbstainShiftGptOssCI} & {\footnotesize\abPartialInsistsAbstainShiftGptOssCI}\\
\bottomrule
\end{tabularx}
\end{table}

\begin{table}[!ht]
\centering\small
\caption{\textbf{How the two search harnesses differ.} Retriever, tools, result count, preview
length, round limit, temperature and model revisions are the same in both.}
\label{tab:harness}
\begin{tabularx}{\textwidth}{@{}>{\raggedright\arraybackslash}p{2.6cm} >{\raggedright\arraybackslash}X >{\raggedright\arraybackslash}X@{}}
\toprule
Setting & Observed runs & Intervention runs\\
\midrule
Thinking & enabled for Nemotron and Gemma; Qwen3.5 as set by its default template & disabled for both models\\
Output cap, non-final calls & 16{,}384 tokens & 4{,}096 tokens\\
Questions and replicates & 100; three repetitions of the wrong-hint contrast & 25 or 26 AI-reviewed questions; one episode per cell\\
Scorer & normalized exact or substring match & strict, line-anchored; bold-wrapped answer lines and unparsable answers count as not correct\\
Arms & no delivered evidence & four evidence arms crossed with the two hint conditions; passages appended to the system message, capped at 12{,}000 characters\\
Hint source & own wrong answer, with a fallback to another question's answer & own recorded wrong answer on the reviewed questions\\
Unaided wrong-hint episodes without a parsable answer & --- & Nemotron \hcFunnelWrongInvalidPctNemotron, Qwen3.5 \hcFunnelWrongInvalidPctQwen\\
\bottomrule
\end{tabularx}
\end{table}

\FloatBarrier
\section{Judge audit}
\label{app:judge}
The supplied-evidence runs label outcomes with an open-weight judge, and strict string matching is the sensitivity analysis. The audit asks how far
judge labels agree with the strict labels already computed for the intervention runs and the curated study. Gemma-4-31B and gpt-oss-120b each relabel
1{,}770 outcomes from those runs, and agreement with the strict labels is summarised by Cohen's kappa and percent agreement
over three classes: correct, names the candidate, and anything else.

Both judges agreed closely with the strict labels. Gpt-oss-120b agreed on \auditAgreeGptOss{} of
the \auditNGptOss{} outcomes (kappa \auditKappaGptOss; \auditKappaCorrectGptOss{} for correct
against not correct and \auditKappaCandidateGptOss{} for naming the candidate), and Gemma-4-31B on
\auditAgreeGemma{} (kappa \auditKappaGemma; \auditKappaCorrectGemma{} and
\auditKappaCandidateGemma). Excluding each judge's own outputs left kappa at
\auditKappaNonSelfGptOss{} and \auditKappaNonSelfGemma. Most disagreements were responses that the
strict scorer could not parse, such as a bold-wrapped answer line, which the judges read as correct
or as naming the candidate; on the \auditParsableNGptOss{} outcomes with a parsable strict label,
agreement was \auditAgreeParsableGptOss{} for both judges (kappa \auditKappaParsableGptOss). The
rest were answers that the strict scorer called other wrong and the judges called correct, mostly the gold
answer stated in another form or with extra words, and in a few cases a paraphrase that a strict
reader might not accept. On the supplied-evidence responses, the
two judges agreed with each other on \studyTwoJudgeAgreementPercent{} of the 20\% sample (kappa
\studyTwoJudgeAgreementKappa). No label was changed by hand.

\section{Questions}
\label{app:questions}

\begin{table}[H]
\centering
\small
\caption{The nine questions in the curated study, by gold answer and the hinted candidate used
for Nemotron (each model received its own recorded wrong answer). For the question with gold
answer 65, the candidate 67 appears in the question text itself.}
\label{tab:questions}
\begin{tabular}{@{}lll@{}}
\toprule
Gold answer & Hinted candidate (Nemotron) & Answer type\\
\midrule
Fort Smith Museum of History & Field Museum of Natural History & institution\\
Sol Campbell & David James & person\\
65 & 67 & number\\
115 & 50 & number\\
Florian Mussgnug & David Luger & person\\
Cory Monteith & Logan Henderson & person\\
Lenny Juma & Edward James Olmos & person\\
Afrigo Band & Lunica & organisation\\
Alexis Nicolas & Jimmy Hunt & person\\
\bottomrule
\end{tabular}
\end{table}

\section{Prompt templates}
\label{app:prompts}

\paragraph{Standard system prompt.}
\begin{quote}\small
Please answer the research question using the information in this conversation and any
experiment-provided evidence where relevant. Do not initiate tools or claim tool use. Give a
concise answer, or state that you cannot determine it. End with exactly one line: Exact answer:
\textless answer\textgreater.
\end{quote}

\paragraph{Careful addition.}
\begin{quote}\small
Think carefully about the question and aim for an accurate, well reasoned response. Review the
information available in this conversation and consider its relevance to the question. Avoid
careless mistakes, stay focused on the task, and present your response clearly and concisely. Use
evidence thoughtfully when formulating the answer.
\end{quote}

\paragraph{Verify-first addition.}
\begin{quote}\small
Treat any proposed answer as a hypothesis. Identify the received evidence most relevant to it, then
determine whether that evidence supports, contradicts, or leaves the hypothesis unresolved. Give the
answer supported by the evidence, or abstain if it is insufficient. Cite only source documents
actually supplied in this conversation.
\end{quote}

\paragraph{Curated-evidence message sequence.}
(1) system prompt; (2) user: ``Research question: \ldots''; (3) wrong-hint arm only, assistant:
``My best guess without searching is \{candidate\}, but I'm not sure.''; (4) evidence block,
``Experiment-provided search-tool result (supplied by the study; not a voluntary tool call by the
model).'' followed by ``Source document: \textit{N}'' and ``Excerpt: \ldots'' for each source; (5)
user: ``Please provide your concise final answer now.'' In the user-message format, step~4 is
joined with adjacent user text. In the system-message format, step~4 is a second system message.

\paragraph{Supplied-evidence message sequence.}
The standard system prompt is followed by the research question. The assistant variants in
\cref{tab:wordings} insert an assistant message followed by a new user message. The
user-insistence variant appends the hint to the question. The evidence block retains the
source and excerpt fields above, followed by the request for a concise final answer. If the
initial response reaches the output limit or lacks a parsable answer, the assistant response is
preserved and the user adds: ``Please stop here and give your final answer now. End with exactly
one line: Exact answer: \textless answer\textgreater.'' This additional turn is capped at 512
tokens. A parsable answer from that turn takes precedence over the initial response.

\paragraph{Supplied-evidence judging.}
At temperature zero, a judge receives the question, benchmark answer and accepted aliases,
candidate answer, and response text. The judge extracts the final answer and selects one
outcome from the four categories in \cref{sec:outcomes}. The prompt omits the subject checkpoint
and experimental condition.

\section{Experimental cohorts and instruction variants}
\label{app:versions}

\begin{table}[H]
\centering
\small
\caption{\textbf{Curated-evidence and parser experiments.} Evidence was inserted as either user
text or a second system message. Counts include unparsable responses.}
\label{tab:runs}
\begin{tabularx}{\textwidth}{@{}l l l >{\raggedright\arraybackslash}X r@{}}
\toprule
Run & Model & Format & Contents & Responses\\
\midrule
Curated & Qwen3.5-35B-A3B & User & 4 arms $\times$ 3 instructions & 216\\
Curated & Nemotron-3 Nano & User & 4 arms $\times$ 3 instructions & 216\\
Curated & Gemma~4 31B & User & 4 arms, standard instruction & 72\\
Curated & gpt-oss-120b & User & 4 arms, standard instruction & 72\\
Curated & Nemotron-3 Nano & System & 4 arms $\times$ 3 instructions & 216\\
Curated & Qwen3-4B (NF4) & System & 4 arms $\times$ 3 instructions & 216\\
Extension & Qwen3.5-35B-A3B & User & 16 questions, 4 arms $\times$ 3 instructions & \qexpResponses\\
Extension & Nemotron-3 Nano & User & 17 questions, 4 arms $\times$ 3 instructions & \expResponses\\
Parser check & Qwen3-4B (NF4) & System & 100 questions, 2 record versions, 2 hint arms, 2 repetitions & 800\\
\bottomrule
\end{tabularx}
\end{table}

\begin{table}[!ht]
\centering\small
\caption{\textbf{Observed search cohorts.} The primary search design comprises 30 conditions
for each of 100 questions for Qwen3.5 and Nemotron. Auxiliary contrasts use available pairs.
For gpt-oss-120b under the primary search settings, the matched question count varies by
contrast and the analysis uses episode-level scores.}
\label{tab:s1pop}
\begin{tabular}{@{}lrr@{}}
\toprule
Run & Questions & Episodes\\
\midrule
Qwen3.5, main design & 100 & 3{,}000\\
Nemotron, main design & 100 & 3{,}000\\
Nemotron, each alternative retriever (BM25, Qwen3-Embedding-8B) & 100 & 500\\
gpt-oss-120b, BM25 & 100 & 500\\
gpt-oss-120b, available pairs & varies by contrast & 1{,}564\\
Nemotron, exploratory sample & 150 & 1{,}800\\
Gemma-4-31B, available pairs & 24 & 750\\
\bottomrule
\end{tabular}
\end{table}

\begin{table}[!ht]
\centering\small
\caption{\textbf{Curated outcomes by instruction.} Each row contains 72 responses, with
18 responses per evidence--hint arm. Mean length is measured in generated tokens.}
\label{tab:instructions}
\setlength{\tabcolsep}{4pt}
\begin{tabular*}{\textwidth}{@{\extracolsep{\fill}}lllrrrrrr@{}}
\toprule
Checkpoint & Role & Instruction & Correct & Adopted & Other wrong & Abstained & Unparsable & Mean tokens\\
\midrule
Nemotron & User & Standard & \nemProtoStdCorrect & \nemProtoStdPlantedcandidate & \nemProtoStdOtherincorrect & \nemProtoStdAbstention & \nemProtoStdInvalid & \nemProtoStdTokens\\
 & & Careful & \nemProtoGenCorrect & \nemProtoGenPlantedcandidate & \nemProtoGenOtherincorrect & \nemProtoGenAbstention & \nemProtoGenInvalid & \nemProtoGenTokens\\
 & & Verify-first & \nemProtoVerCorrect & \nemProtoVerPlantedcandidate & \nemProtoVerOtherincorrect & \nemProtoVerAbstention & \nemProtoVerInvalid & \nemProtoVerTokens\\
\addlinespace
Qwen3.5 & User & Standard & \qwenProtoStdCorrect & \qwenProtoStdPlantedcandidate & \qwenProtoStdOtherincorrect & \qwenProtoStdAbstention & \qwenProtoStdInvalid & \qwenProtoStdTokens\\
 & & Careful & \qwenProtoGenCorrect & \qwenProtoGenPlantedcandidate & \qwenProtoGenOtherincorrect & \qwenProtoGenAbstention & \qwenProtoGenInvalid & \qwenProtoGenTokens\\
 & & Verify-first & \qwenProtoVerCorrect & \qwenProtoVerPlantedcandidate & \qwenProtoVerOtherincorrect & \qwenProtoVerAbstention & \qwenProtoVerInvalid & \qwenProtoVerTokens\\
\addlinespace
Qwen3-4B & System & Standard & \qwthreeProtoStdCorrect & \qwthreeProtoStdPlantedcandidate & \qwthreeProtoStdOtherincorrect & \qwthreeProtoStdAbstention & \qwthreeProtoStdInvalid & \qwthreeProtoStdTokens\\
 & & Careful & \qwthreeProtoGenCorrect & \qwthreeProtoGenPlantedcandidate & \qwthreeProtoGenOtherincorrect & \qwthreeProtoGenAbstention & \qwthreeProtoGenInvalid & \qwthreeProtoGenTokens\\
 & & Verify-first & \qwthreeProtoVerCorrect & \qwthreeProtoVerPlantedcandidate & \qwthreeProtoVerOtherincorrect & \qwthreeProtoVerAbstention & \qwthreeProtoVerInvalid & \qwthreeProtoVerTokens\\
\addlinespace
Nemotron & System & Standard & 24 & 4 & 40 & 3 & 1 & ---\\
 & & Careful & 30 & 6 & 28 & 5 & 3 & ---\\
 & & Verify-first & 13 & 1 & 32 & 14 & 12 & ---\\
\bottomrule
\end{tabular*}
\end{table}

\begin{table}[!ht]
\centering\small
\caption{\textbf{Nemotron with evidence in system messages.} Responses under the standard
instruction (18 per arm). Under the wrong hint, the 95\% interval for the adoption difference
between decisive and neutral evidence is \nemoneDOneCI{} percentage points
(50{,}000 bootstrap draws).}
\label{tab:system-message-outcomes}
\begin{tabular}{@{}llrrrrr@{}}
\toprule
Hint & Evidence & Correct & Adopted & Other wrong & Abstained & Unparsable\\
\midrule
None & Neutral & \nemoneNoHintNeutralCorrect & \nemoneNoHintNeutralPlanted & \nemoneNoHintNeutralOther & \nemoneNoHintNeutralAbstain & \nemoneNoHintNeutralInvalid\\
Wrong & Neutral & \nemoneWrongNeutralCorrect & \nemoneWrongNeutralPlanted & \nemoneWrongNeutralOther & \nemoneWrongNeutralAbstain & \nemoneWrongNeutralInvalid\\
None & Decisive & \nemoneNoHintDecisiveCorrect & \nemoneNoHintDecisivePlanted & \nemoneNoHintDecisiveOther & \nemoneNoHintDecisiveAbstain & \nemoneNoHintDecisiveInvalid\\
Wrong & Decisive & \nemoneWrongDecisiveCorrect & \nemoneWrongDecisivePlanted & \nemoneWrongDecisiveOther & \nemoneWrongDecisiveAbstain & \nemoneWrongDecisiveInvalid\\
\bottomrule
\end{tabular}
\end{table}

\FloatBarrier
\section{Compute}
\label{app:compute}
The supplied-evidence and search experiments used NVIDIA H100 GPUs on the IIT Kharagpur
DGX AI Cluster. The curated Qwen3-4B experiments and synthetic parser check used NVIDIA V100
GPUs on PARAM Shakti at IIT Kharagpur.

\end{document}